\documentclass[twocolumn,aps,pra,showpacs,raggedbottom,nobalancelastpage,amssymb,superscriptaddress]{revtex4-1}

\usepackage{graphicx}
\usepackage{amsmath}
\usepackage{amssymb}
\usepackage{bm}
\usepackage[usenames]{color}
\usepackage{amsfonts}
\usepackage{appendix}
\usepackage{dsfont}
\usepackage{float}

\definecolor{Nathanblue}{rgb}{0.,0.24,0.51}

\def\be{\begin{equation}}
\def\ee{\end{equation}}

\def\bs#1{\boldsymbol{#1}}

\usepackage{amssymb}
\usepackage{mathrsfs}
\usepackage{graphicx}
\usepackage{dcolumn}
\usepackage{bm}
\usepackage{amsmath}
\usepackage{amsfonts}
\usepackage{color}
\usepackage{float}
\usepackage{graphicx}
\usepackage{dcolumn}
\usepackage{bm}
\usepackage{hyperref}
\usepackage{xcolor}
\hypersetup{
	colorlinks,
	linkcolor={red!50!black},
	citecolor={green!50!black},
	urlcolor={blue!80!black}
}
\usepackage[caption=false]{subfig}
\makeatletter
\newcommand{\colorcaption}[2][]{%
  \begingroup%
  \renewcommand{\@caption@fignum@sep}{ (color online). }%
  \caption[#1]{#2}%
  \endgroup%
}
\makeatother

\begin{document}

\preprint{APS/123-QED}
\title{Electromagnetic response of quantum Hall systems in dimensions five and six and beyond}

\author{Ching Hua Lee}
\email{calvin-lee@ihpc.a-star.edu.sg}
\affiliation{Institute of High Performance Computing, 138632, Singapore}
\affiliation{Department of Physics, National University of Singapore, 117542, Singapore}

\author{Yuzhu Wang }
\affiliation{School of Physics, Sun Yat-sen University, Guangzhou 510275, China}

\author{Youjian Chen}
\affiliation{Department of physics and astronomy,Stony Brook University ,Stony Brook,NY 11794-3800}

\author{Xiao Zhang}
\email{zhangxiao@mail.sysu.edu.cn}
\affiliation{School of Physics, Sun Yat-sen University, Guangzhou 510275, China}

\date{\today}

\begin{abstract}
Quantum Hall (QH) states are arguably the most ubiquitous examples of nontrivial topological order, requiring no special symmetry and elegantly characterized by the first Chern number. Their higher dimension generalizations are particularly interesting from both mathematical and phenomenological perspectives, and have attracted recent attention due to high profile experimental realizations~\cite{Lohse2017Exploring,zilberberg2018photonic}. In this work, we derive from first principles the electromagnetic response of QH systems in arbitrary number of dimensions, and elaborate on the crucial roles played by their modified phase space density of states under the simultaneous presence of magnetic field and Berry curvature. We provide new mathematical results relating this phase space modification to the non-commutativity of phase space, and show how they are manifested as a Hall conductivity quantized by a higher Chern number. When a Fermi surface is present, additional response currents unrelated to these Chern numbers also appear. This unconventional response can be directly investigated through a few minimal models with specially chosen fluxes. These models, together with more generic
6D QH systems, can be realized in realistic 3D experimental setups like cold atom systems through possibly entangled synthetic dimensions.
\end{abstract}
\maketitle

\section{\label{sec:level1}Introduction}

The discovery of quantum Hall (QH) effects in 2D electron gases in 1980 and Graphene in 2005 were among the greatest highlights of modern condensed matter physics~\cite{von1986quantized,laughlin1981quantized,jain1989composite,zhang2005experimental,gusynin2005unconventional,novoselov2007room}. They sparked of the era of topologically protected order, where topological invariants lead to exactly quantized measurable quantities like Hall conductivity~\cite{bernevig2006quantum,zhang2009topological,qi2010quantum,qi2011topological,hasan2010colloquium,zhang2012actinide,He03052016}. Indeed, in the last decade, lattice analogs of QH fluids known as Chern or quantum anomalous Hall insulators have taken central stage in theoretical~\cite{liu2008quantum,yu2010quantized,sheng2011fractional,ezawa2012valley,yao2013realizing,bergholtz2013topological,wang2013anomalous,lee2014lattice,claassen2015position,lee2017band,Wu2008Orbital} and experimental circles~\cite{chang2013experimental,chang2015high,Zhang2011Quantum,Qiao2010Quantum,Zhang2012Electrically,Liu2010Quantum,Tse2011Quantum,Jiang2012Quantum,Zhang2014Quantum,Wang2014Quantum}, with their dissipationless edge states holding enticing technological promises~\cite{zhang2010topological,Yao2015Ultra,Chen2012Photomixing,Wang2009Broadband,Zheng2017Group,Xu2015Ultrasensitive,Hor2009p,Shao2015Photoconductivity,Zhang2017Topological}. More recently, Chern insulators have also played pivotal roles in theoretical investigations and experimental realizations of chiral Majorana edge modes in proximitized topological superconductors~\cite{he2017chiral,beenakker2015random,lian2017non,chan2017non,yap2017photoinduced}.

While intrinsically 2D phases, QH states can, through special choices of gauge, also be expressed as independent 1D quasiperiodic chains labeled by a synthetic dimensional parameter. This fortuitous dimensional reducibility has allowed for realistic 2D implementations of higher dimensional QH generalizations~\cite{kraus2013four,price2015four,lu2016topological,Ozawa2016Synthetic,4DphotonicQH}: recent experiments~\cite{Lohse2017Exploring,zilberberg2018photonic} have demonstrated signatures of the 4D QH effect with ultracold bosonic atoms in optical superlattices and tunable photonic waveguides. In these setups, a $d$-dimensional QH system was constructed simply by taking the tensor product of $\frac{d}{2}$ copies of 2D QH systems~\cite{price2015four}, with the 
resultant $d$-dimensional magnetic flux decomposable into a product of ordinary 2D fluxes. One can alternatively construct nontrivially ``entangled'' higher-dimensional QH states by introducing additional off-diagonal couplings between the physical and synthetic dimensions~\cite{4DphotonicQH}, as shall be detailed in Sect.\ref{sec:entangled}. So far, existing proposals and experimental realizations have been limited to 4D QH phases, although QH phases in 5 or more dimensions are just as realizable in principle~\cite{Lian2016Five,Lian2017Weyl}.

In all time-reversal broken systems including QH fluids and Chern insulators, small changes in the magnetic field lead to non-perturbative modifications of the lattice band structure due to non-commensurability effects, as epitomized by the Hofstadter butterfly~\cite{Hofstadter1993Energy}. Yet, their \emph{physical} electromagnetic responses are of course only perturbatively modified~\cite{xiao2006berry,qi2008topological,lee2015negative,bulmash2015unified,price2015four,morimoto2016semiclassical}. This suggests the strategy of separating a generic time-reversal breaking flux into an ``intrinsic'' Berry curvature that redefines the band structure, and an ``external'' magnetic field that only enters the physical response as a perturbing field~\cite{price2015four,price2016measurement}. At the level of semiclassical response dynamics, this simultaneous presence of magnetic field and Berry curvature seemingly leads to a subtle but interesting violation of Liouville's theorem, which can be remedied either by redefining the phase space density of states~\cite{xiao2005berry}, or by resorting to non-canonical coordinates~\cite{duval2006berry,gosselin2006semiclassical}. Both of these resolutions shall give rise to especially profound physical consequences in higher dimensions, as we will detail in this work. Specifically, the interplay of magnetic field and Berry curvature not only produces a quantized third Chern number Hall response in 6 or more dimensions, but also yields new current responses when a Fermi surface is present.

Our paper is structured as follows. In Sect. II, we first derive the semiclassical equations of motion (EOMs) of electronic wavepackets in the presence of both magnetic field and Berry curvature, starting from their Lagrangian. Next we derive new explicit results for the concomitant modification of the phase space density of states in arbitrary number of dimensions. Combining these results, we present the current response terms and re-derive the latter in terms of topological pumping. Following that in Sect. III, we present minimal models for studying interesting non-topological surface current contributions in 5 or 6 dimensions when a Fermi surface is present. We next discuss how these 6D QH systems can be realized with 3D physical systems with 3 synthetic parameters in Sect. IV, before reviewing specifics of their experimental realization in cold atom systems in Sect.~V.

\section{Semiclassical transport of electronic bands with nonzero Berry curvature and flux}

We consider generic $d$-dimensional electronic systems subject to external perturbing electric and magnetic fields $\bs E(\bs r) = E_{\mu}(\bs r) \bs{e}^{\mu}=-\nabla_\mu\Phi(\bs r)\,\bs{e}^{\mu}$ and $B_{\mu \nu}(\bs r)  = \partial_{\mu}A_{\nu}(\bs r)- \partial_{\nu}A_{\mu}(\bs r) $, where $\Phi(\bs r)$ and $\bs A(\bs r)  = A_{\mu}(\bs r) {\bs e}^{\mu}$ are the scalar and vector potentials respectively. In dimensions $d>3$, the magnetic field cannot be visualized as a vector, but is rather a 2nd-rank tensor field with $d(d-1)/2$ independent components. Introducing a lattice regularization if necessary, the energy eigenstates of the system splits into distinct bands on the momentum-space d-torus $\mathbb{T}^d$. When intrinsic time-reversal or inversion/particle-hole asymmetry is present, each band $|\psi\rangle$ will also acquire a Berry curvature $\Omega^{\mu\nu}(\bold k)=\partial^{\mu}a^{\nu}(\bs k)- \partial^{\nu}a^{\mu}(\bs k) =i(\langle \partial^{k_\mu}\psi|\partial^{k_\nu}\psi\rangle-\langle \partial^{k_\nu}\psi|\partial^{k_\mu}\psi\rangle)$, where $a^\mu = -i\langle\psi|\partial^{k_\mu} \psi\rangle$ is the Berry connection.

In this work, we shall assume non-degenerate bands, such that each band (partially or completed filled) can be treated separately. The current response $\bs j =j^\mu {\bs e}_{\mu}$ is given by
\begin{equation}
j^\mu = \int_{\mathbb{T}^d}[\dot r^\mu D(\bs r,\bs k)]d^dk,
\label{jmu}
\end{equation}
where the electromagnetic fields $\bs E$, $B_{\mu\nu}$ and the Berry curvature $\Omega^{\mu\nu}$ implicitly affect the quantities in the square parentheses. In the following subsections, we shall detail how exactly they affect the semiclassical wavepacket velocity $\dot r^\mu$, as well as the phase space density of states $D(\bs r,\bs k)$. The former will be described at the level of classical equations of motion in Sect.~\ref{sec:EOM}, and recast in terms of topological pumping in Sect.~\ref{sec:pumping}.

\subsection{Semiclassical equations of motion}
\label{sec:EOM}

We first attempt to understand the response properties of an electronic system from the equations of motion of its semi-classical electron wavepackets. We consider wavepackets which are much larger than the lattice spacing ($\sim 1 \AA$), such that each of them possesses a single well-defined peak in momentum space. The wavepackets must also be such smaller than spatial variation of the magnetic vector potential $\bs A(\vec r)$, which hence restricts the external magnetic field to be a weak perturbation. The external electric field must also be sufficiently weak for the wave-packet to evolve adiabatically without interband transitions.

Labeling the spatial and momentum-space peaks of a wavepacket as $\bs r=r^{\mu} {\bs e}_{\mu}$ and $\bs k=k_{\mu} {\bs e}^{\mu}$, one can write down the Lagrangian (with $e$ and $\hbar$ set to unity)
\begin{eqnarray}
&&L=k_\mu \dot r^\mu+ \dot k_\mu a^\mu(\bs k)-\dot r^\mu A_\mu(\bs r) - H(\bs r,\bs k),\notag\\
&&\notag\\
&&H(\bs r,\bs k)=\mathcal{E}(\bs k)-\Phi(\bs r).
\label{lagrangian}
\end{eqnarray}
through minimal coupling of both the magnetic and Berry gauge fields $A_\mu(\bs r)$ and $a^\mu(\bs k)$, with the Hamiltonian $H(\bs r,\bs k)$ simply given by the sum of the Bloch Hamiltonian $\mathcal{E}(\bs k)$ and the electrical potential $-\Phi(\bs r)$. Solving the Euler-Lagrange equations with Eq.~\ref{lagrangian}, we obtain the equations of motion (EOMs) for the wavepacket\footnote{But see Ref.~\cite{marder2010condensed} for an alternative pedagogical derivation} (Einstein summation implied from now on)
\begin{align}
\dot{r}^\mu  &= v^\mu(\bs k) - \dot{k}_\nu \Omega^{\mu \nu}  (\bs k), \label{eq:semir} \\
\dot{k}_\mu &= - E_\mu(\bs r) - \dot{r}^\nu  B_{\mu \nu}(\bs r) \,, \label{eq:semik}
\end{align}
where $v^\mu=\frac{\partial \mathcal{E} (\bs k)}{\partial k_\mu }$ is the group velocity computed from the wavepacket energy of the unperturbed Bloch band. In the above, the momentum and positional peak positions take on symmetrical roles, being linearly accelerated by position-momentum duals $v^\mu$ and $-E_\mu$, and ``curved'' by analogously dual fields $\Omega^{\mu\nu}$ and $B_{\mu\nu}$. This position-momentum symmetric formulation pertains to wavepackets in \emph{any} number of dimensions $d$, and is one of many examples of position-momentum duality. In related contexts, position-momentum symmetry has been employed to prove results in the free-fermion entanglement spectra of spin textures and thermalized systems~\cite{lee2014position,lai2015entanglement}. Very prevalently, it has also allowed an elegant formulation of 2D QH dynamics in terms of cyclotron and guiding-center coordinates~\cite{prange1982conduction,cary2009hamiltonian}, which has incidentally been exploited for relating fractional QH phases in lattice and continuum systems~\cite{claassen2015position}.

By repeatedly substituting Eq.~\ref{eq:semik} into Eq.~\ref{eq:semir}, we can obtain the equation of motion for the center-of-mass (CM) velocity $\dot{\bs r}$ up to any desired level of accuracy. For instance, up to the third order in $E_\mu$ and $B_{\mu\nu}$,
\begin{widetext}
\begin{align}
\dot{r}^\mu &= v^\mu+  E_\nu \Omega^{\mu \nu}+  \dot{r}^\gamma  B_{ \nu \gamma} \Omega^{\mu \nu} \nonumber\\
&= v^\mu+  E_\nu  \Omega^{\mu \nu}+ \left( v^\gamma +  E_\delta  \Omega^{\gamma \delta} +  \dot{r}^\alpha  B_{ \delta \alpha}  \Omega^{\gamma \delta}
\right)B_{ \nu \gamma}  \Omega^{\mu \nu}
\nonumber \\
&\approx v^\mu+  E_\nu  \Omega^{\mu \nu}+ \left( v^\gamma+  E_\delta  \Omega^{\gamma \delta} +  \left (v^\alpha +  E_\beta \Omega^{\alpha \beta}  +v^\theta  B_{ \beta \theta} \Omega^{\alpha \beta} \right) B_{ \delta \alpha}  \Omega^{\gamma \delta} \right)B_{ \nu \gamma}  \Omega^{\mu \nu}
\label{EOM}
\end{align}
\end{widetext}
Since our eventual purpose is to obtain only the leading-order terms of each type of transport contribution, we have neglected field-induced shifts in the wavepacket CM, which can be expressed in terms of the quantum geometric metric as in Ref.~\cite{Gao2014Field}.

\subsection{Modification of density of states}
\label{sec:DOS}
A central foundation of Hamiltonian mechanics is Liouville's theorem, which states that the phase space volume must remain conserved under time evolution, i.e. that $\frac{d(\ln \Delta V)}{dt}=\nabla_{\bs r}\cdot \dot{ \bs r} + \nabla_{\bs k}\cdot \dot{ \bs k}$ should vanish for a suitably defined volume form $\Delta V$. With the EOMs in Eqs.~\ref{eq:semir} and \ref{eq:semik}, it is easy to show, at least in 2D~\cite{xiao2005berry,duval2006berry}, that the simultaneous existence of Berry curvature and magnetic field lead to non-conservation of the naively defined volume element $\Delta V_{naive} =\Delta \bs r \Delta \bs k$. To derive the properly conserved volume element and remedy this apparent conundrum, we rewrite the EOMs Eqs.~\ref{eq:semir} and \ref{eq:semik} in the language of Hamiltonian dynamics:
\begin{equation}
\dot r^\mu = \{ H,r^\mu\},\;\;\;\; \dot k_\mu=\{ H,k_\mu\},
\label{PB}
\end{equation}
where $\{ u,v\}$ is the Poisson bracket defined by $\{ u,v\}=\omega^{\mu\nu}\partial_\mu u \,\partial_\nu v$. $\omega^{\mu\nu}$ is the inverse of the matrix $\omega_{\mu\nu}$, which defines the closed symplectric structure of this Hamiltonian system, and famously gives the volume element
\begin{equation}
\Delta V = \text{Pf}(\omega_{\mu\nu})\Delta \bs r \Delta \bs k
\end{equation}
that will indeed remain conserved under time evolution~\cite{duval2006berry}. $\text{Pf}$ denotes the Pfaffian, the square root of the determinant operator on an even-dimensional antisymmetric matrix. To compute the modification factor $\text{Pf}(\omega_{\mu\nu})$, we contract Eq.~\ref{PB} with $\omega_{\mu\nu}=(\omega^{\mu\nu})^{-1}$ and obtain $\omega_{\mu\nu}$ as the real antisymmetric transformation matrix between $(\dot{\bs r},\dot{\bs k})$ and $(\bs E,\bs v)$, i.e.
\begin{equation}
\omega_{\mu\nu}=\left(\begin{matrix}
  -B & -\mathbb{I}_{d\times d} & \\
\mathbb{I}_{d\times d} & \Omega &
\end{matrix}\right)
\label{omegamunu}
\end{equation}
where $  B$ and $ \Omega$ are the magnetic and Berry curvature tensors respectively. With a detailed derivation left to Appendix \ref{sec:det}, we obtain our main result on the modification to the density of states:
\begin{widetext}
\begin{eqnarray}
D(\bs r,\bs k)&=&\frac1{(2\pi)^d}\text{Pf}(\omega_{\mu\nu})
=\frac1{(2\pi)^d} \sqrt{\text{det}(\mathbb{I}_{d\times d}-  B \Omega)}\notag\\
&=&\frac1{(2\pi)^d}\left[1+\epsilon^{\mu_1\mu_2...\mu_d}\varepsilon_{\nu_1\nu_2...\nu_d}\sum_{j>0}\frac1{4^{2j-1}(d-2j)!}\left(\prod_{l=1}^j B_{\mu_{2l-1}\mu_{2l}}\Omega^{\nu_{2l-1}\nu_{2l}}\right)\prod_{l'=2j+1}^d\delta_{\mu_{l'},\nu_{l'}}\right],
\label{Drk}
\end{eqnarray}
\end{widetext}
where $\varepsilon^{\mu_1\mu_2...\mu_d},\varepsilon_{\mu_1\mu_2...\mu_d}$ are generalized Levi-Civita symbols which are
equal to the sign of the permutation $\{\mu_1,...,\mu_d\}$ for unique $\mu_j$ indices, and vanish if there is any repeated $\mu_j$ indices. As such, for a given number of dimensions $d$, only $2j$-th order terms with $j\leq \lfloor\frac{d}{2}\rfloor$ exist. Eq.~\ref{Drk} generalizes previous results~\cite{xiao2005berry,price2015four} to arbitrary number of dimensions $d$, with expressions for the first several $d$ given by
\begin{align}
D_{d=\text{2D/3D}}=&  \frac{1}{(2\pi)^d} \left[1 + \frac{1}{2} B_{\mu \nu} \Omega^{\mu \nu}\right] \,,\\
D_{d=4D/5D} = & \frac{1}{(2\pi)^d} [  1 + \frac{1}{2}  B_{\mu \nu} \Omega^{\mu \nu}   +\frac{1}{64} \left( \varepsilon^{\alpha \beta \gamma \delta } B_{\alpha \beta }B_{\gamma \delta }\right)  \nonumber \\
&  \times \left( \varepsilon_{\mu \nu \lambda \rho} \Omega^{\mu \nu }\Omega^{\lambda \rho }\right) ] ,
\label{D4D}
\end{align}
\begin{align}
D_{6D/7D} = & \frac{1}{(2\pi)^d} [  1 + \frac{1}{2}  B_{\mu \nu} \Omega^{\mu \nu}   +\frac{1}{64} \left( \varepsilon^{\alpha \beta \gamma \delta } B_{\alpha \beta }B_{\gamma \delta }\right)  \nonumber \\
&  \times \left( \varepsilon_{\mu \nu \lambda \rho} {\Omega}^{\mu \nu }\Omega^{\lambda \rho }\right) + \frac{1}{1024} \left( \varepsilon^{\zeta \eta \theta \tau \kappa \xi } B_{\zeta \eta }B_{\theta \tau }B_{\kappa \xi }  \right) \nonumber \\
&\times   \left( \varepsilon_{\sigma \omega \iota \phi \chi \psi } {\Omega}^{\sigma \omega }{\Omega}^{\iota \phi }{\Omega}^{\chi \psi }\right) ] \,,
\end{align}
In the above, there is an implicit sum over all possible subsets of indices whenever the there are more dimensions than indices, i.e. the term containing $\varepsilon^{\alpha\beta\gamma\delta}$ will be summed over all 4-element subsets $\alpha,\beta,\gamma,\delta\in\{1,2,3,4,5\}$ in the 5D case. This is consistent with the physical intuition that additional synthetic dimensions should only result in new response terms, and not modify existing terms.

\subsection{Leading order current response terms }
Having discussed the semiclassical EOMs and how they crucially modify the density of states, we are now ready to present results for the electromagnetic response.

In four or fewer dimensions, it is well-established~\cite{price2015four} that the response of an insulating electronic system (with completed filled band) is completely quantized, with a Hall response term proportional to the 1st Chern number, and a magneto-electric response term proportional to the 2nd Chern number. Analogous results hold in higher dimensions. Specializing to six dimensions in anticipation of the physically realizations discussed later, a 6D insulating system has a bulk response current:
\begin{align}
j^\mu=j^\mu_{bulk}&= \int_{\mathbb{T}^6} { \text{d}^6 k } [ \dot{{ r}}^\mu D({\bs r}, {\bs k})]\notag\\
   &= \frac{\mathcal{C}^{\mu\nu}_1}{2\pi}E_\nu  +  \frac{1}{2} \frac{\mathcal{C}^{\mu\alpha\beta\nu}_2}{(2 \pi)^2}  B_{ \alpha \beta}E_\nu   \nonumber \\
 &+ \frac{1}{8} \frac{\mathcal{C}_3}{\left(2\pi\right)^3}  \varepsilon^{\mu \alpha \beta  \delta \gamma \nu}   B_{ \alpha \beta} B_{\delta \gamma} E_\nu 
\label{eq:curr}
\end{align}
In the above, $\mathcal{C}_1^{\mu\nu}$, $\mathcal C_2^{\mu\alpha\beta\nu}$ and the topologically quantized third Chern number $\mathcal{C}_3$ are given by
\begin{align}
\mathcal{C}^{\mu\nu}_1\!&=\!\frac{1}{(2 \pi)^5} \int_{\mathbb{T}^6}
 {\Omega}^{\mu \nu}   \text{d}^6 k\\
\mathcal{C}^{\mu\alpha\beta\nu}_2\!
&=\!\frac{1}{(2 \pi)^4} \int_{\mathbb{T}^6}  \left( \Omega^{\mu \alpha}\Omega^{\beta \nu}\!-\! \Omega^{\mu\nu }\Omega^{\beta \alpha}\!+\! \Omega^{\beta \mu}\Omega^{\alpha \nu} \right) \text{d}^6k  \label{second_chern}
\end{align}
\begin{align}
 \mathcal{C}_3\!&=\!\frac{1}{3! \times (2\pi)^3} \int_{\mathbb{T}^6}  \Omega\wedge\Omega\wedge\Omega \nonumber \\
& = \frac{1}{8\pi^3} \int_{\mathbb{T}^6} \left(\Omega^{zs} \Omega^{wt} \Omega^{xy} \right.
+ \Omega^{wt} \Omega^{xs} \Omega^{yz}
- \Omega^{zs} \Omega^{xw} \Omega^{yt} \nonumber \\
&\;\;\;\;+ \Omega^{zw} \Omega^{xs} \Omega^{yt}
+ \Omega^{zs}\Omega^{xt} \Omega^{yw}
- \Omega^{zt} \Omega^{xs} \Omega^{yw} \nonumber \\
&\;\;\;\;- \Omega^{wt} \Omega^{xz} \Omega^{ys}
- \Omega^{zw} \Omega^{xt} \Omega^{ys}
+\Omega^{zt} \Omega^{xw} \Omega^{ys} \nonumber \\
&\;\;\;\;- \Omega^{zw} \Omega^{xy} \Omega^{st}
- \Omega^{xw} \Omega^{yz} \Omega^{st}
+ \Omega^{xz} \Omega^{yw} \Omega^{st} \nonumber \\
&\;\;\;\;+ \left.\Omega^{zt} \Omega^{xy} \Omega^{sw}
+ \Omega^{xt} \Omega^{yz} \Omega^{sw}
- \Omega^{xz} \Omega^{yt} \Omega^{sw} \right) \text{d}^6 k
\label{third_chern}
\end{align}
Note that $\mathcal{C}_1^{\mu\nu}$ and $\mathcal C_2^{\mu\alpha\beta\nu}$ are computed with respect to $\{\mu,\nu\}$ and $\{\mu,\alpha,\beta,\nu\}$ hyperplane subsets of the 6D space respectively. Eq.~\ref{eq:curr} can be put into explicit covariant form if we were to explicitly write it in terms of the generalized Levi-Civita symbols.

In the metallic case (with a Fermi surface $\partial\Gamma$), it is known for
$d\leq 4$ that there exists a distinct surface term producing a non-topological surface current\footnote{In the insulating case, these terms will disappear completely due to the Bianchi identity, as in Ref.~\onlinecite{price2015four}.} $\bs j_{surf}$, in addition to corrections to the Chern number contributions from Eq. \ref{eq:curr}. For five or more dimensions, this surface current takes a much more sophisticated form. 
Firstly, $j^\mu_{bulk}$
ceases to be topologically quantized, with its integration region bounded by $\partial\Gamma$. Importantly, the total response current $j^\mu=\left(j^\mu_{bulk}+j^\mu_{surf}\right)|_{\mathbb{T}^6\rightarrow \Gamma}$ now contains a new contribution$j^\mu_{surf}$ given by 
\begin{align}
j_{surf}^{\mu}=&\int_{\Gamma} \frac{ \text{d}^6 k }{(2\pi)^6}    \left(  v_\gamma   B_{ \nu \gamma} \Omega^{\mu \nu} + \frac{1}{2} v_\mu B_{\gamma \nu}  \Omega^{\gamma \nu} \right) \notag\\
&+\int_{\Gamma} \frac{ \text{d}^6 k }{(2\pi)^6}
  \left[ \left( v_\alpha  B_{ \delta \alpha} \Omega^{\gamma \delta}
+ \frac{1}{2}v_\gamma   B_{\delta \alpha}  {\Omega}^{\delta \alpha}  \right) B_{ \nu \gamma} \Omega^{\mu \nu}\right. \nonumber \\
& + \left.
  \frac{1}{64}  v_\mu ( \varepsilon^{\alpha \beta \gamma \delta } B_{\alpha \beta }B_{\gamma \delta} )\times( \varepsilon_{\xi \nu \lambda \rho} {{\Omega}}^{\xi \nu }{\Omega}^{ \lambda \rho} )   \right]
\label{jnon}
\end{align}

As will be explicated in Appendix \ref{sec:derivation}, both the surface current $\bs j_{surf}$ and the topological 3rd Chern number $\mathcal{C}_3$ arise from a careful combination of EOM and density of states modifications (Eqs.~\ref{EOM} and~\ref{Drk}), though with the $l$-th order correction entering only the $\bs j_{surf}$ in $2l+1$ dimensions. Note that Eq.~\ref{jnon} includes only leading (quadratic) order contributions in the magnetic field; we leave the study of its higher order corrections to future work.

\subsection{Semiclassical topological pumping}
\label{sec:pumping}

After examining the wavepacket equations of motion and deriving its semiclassical transport current, which in general does not have to be parallel to the applied fields, it is instructive to re-derive these results for some lower dimensions in terms of topological (Thouless-like) pumping arguments. For the sake of clarity, in this subsection we shall adopt somewhat different notation from the rest of this paper.

Consider the motion of a Bloch wavepacket centered around momentum $\vec k$ and real-space coordinate $X_{\vec k}$ under the influence of a Berry curvature $\Omega$. Semiclassical theory tells us that $X_{\vec k}$ responds to an external electric field $\bs E=E_s\bs e^s$ according to $\dot X_{\vec k}=-\dot  k \times \Omega$, where $ k_s \rightarrow k_s + E_st$ from minimal coupling. Hence, we have
\begin{equation}
X_{\vec k}^s(t)=(k_s+E_st)\Omega_{xs}= X^s(0)+\Omega_{xs}E_st
\end{equation}
where the superscript $s$ has been added to emphasize that the time dependence of $X$ is due to an electric field in the direction $\hat s$. This is just the Hall response of wavepackets.

\subsubsection{Spectral flow for 2D QH}
We next generalize the above Hall response to spectral flow due to topological pumping. For a 2D QH system in the $x$-$s$ plane, the spectral flow of Wannier function centers is directly related to the Hall pumping of its constituent wavepackets. We write the time-dependent Wannier center coordinate as $X(t)$, which is equal to the Wannier polarization $P(k_s)$ upon the minimal coupling $k_s\rightarrow k_s + E_st$:
\begin{eqnarray}
X(t)&=&  P_X(k_s+E_st)\notag\\
&= & \frac1{2\pi}\int_0^{2\pi}\int_0^{k_s+E_st} \Omega_{xs}(k'_x,k'_s)dk'_sdk'_x \notag\\
&\sim & (k_s+E_st)\frac{1}{(2\pi)^2}\int_0^{2\pi}\int_0^{2\pi}\Omega_{xs}(k'_x,k'_s)dk'_sdk'_x \notag\\
&=& (k_s+E_st)\int_{[0,2\pi]^2}\frac{\Omega_{xs}}{(2\pi)^2}d^2\vec k' \notag\\
&=& X(0)+\frac{C_1^{xs}}{2\pi}E_st\notag\\
&=&\frac1{(2\pi)^2}\int_{[0,2\pi]^2}X^s_{\vec k'}(t)d^2\vec k'
\label{2D}
\end{eqnarray}
In the 2nd line, I have used the standard expression for the Wannier polarization~\cite{qi2011generic,huang2012entanglement,lee2013pseudopotential} $P_X$ of $X(t)$ due to $k_s$, which for macroscopic displacements can be linearly approximated as in line 3. 
In the second last line, I have equivalently expressed $X(t)-X(0)$ in terms of the Chern number $C_1^{xs}$.

Not surprisingly, the Wannier center $X(t)$ is given, in the last line, by the integral of the centers of all wavepackets in the BZ.

\subsubsection{Spectral flow for 4D QH}
We next illustrate how to generalize this pumping argument to 4 dimensions.  Notably, we shall see that the modification of the density of states must be included to recover the correct response behavior given by the Chern numbers.

With 4 dimensions, there exists nontrivial pumping contributions up to 2nd order in the perturbing fields $E$ and $B$. Let the 4D QH system span coordinates $x,y,s$ and $u$. 
Qualitatively, we know that the combination of electric field $E_u$ and magnetic field $B_{ys}$ can lead to the pumping of $X(t)$ through three possible mechanisms:
\begin{enumerate}
\item Direct Hall response via $k_u\rightarrow k_u+E_ut$,
\item Second order Hall responses mediated by the magnetic field via $k_s\rightarrow k_s+B_{sy}Y(t)$ and $k_y\rightarrow k_y+B_{ys}S(t)$, where $Y(s)$ and $S(t)$ are themselves pumped by $E_u$, and
\item Second order corrections from phase space modification due to the magnetic field.
\end{enumerate}
Explicitly, we have
\begin{widetext}
\begin{eqnarray}
X(t)&=& \int\frac{dX(t)}{dt}dt\notag\\
&=&\frac1{(2\pi)^4}\int_{[0,2\pi]^4} \int[\dot X^u_{\vec k'}(t)+\dot X^s_{\vec k'}(t)+\dot X^y_{\vec k'}(t)]dt\, d^4\vec k'\notag\\
&=&\frac1{(2\pi)^4}\int_{[0,2\pi]^4} [X^u_{\vec k'}(t)+X^s_{\vec k'}(t)+X^y_{\vec k'}(t)]d^4\vec k'\notag\\
&=&\frac1{(2\pi)^4}\int_{[0,2\pi]^4} [(k_u+E_ut)\Omega_{xu}+(k_s+B_{sy}Y^u_{\vec k'}(t))\Omega_{xs}+(k_y+B_{ys}S^u_{\vec k'}(t))\Omega_{xy}]d^4\vec k'\notag\\
&=&\frac1{(2\pi)^4}\int_{[0,2\pi]^4} [(k_u+E_ut)\Omega_{xu}+(k_s+B_{sy}(k_u+E_ut)\Omega_{yu})\Omega_{xs}+(k_y+B_{ys}(k_u+E_ut)\Omega_{su})\Omega_{xy}]d^4\vec k'\notag\\
&=&X(0)+\frac{t}{(2\pi)^4}\int_{[0,2\pi]^4} [E_u\Omega_{xu}+B_{sy}E_u\Omega_{yu}\Omega_{xs}+B_{ys}E_u\Omega_{su}\Omega_{xy}]d^4\vec k'\notag\\
&=&X(0)+\frac{E_ut}{(2\pi)^4}\int_{[0,2\pi]^4} [\Omega_{xu}+B_{sy}\Omega_{yu}\Omega_{xs}+B_{ys}\Omega_{su}\Omega_{xy}]\left[D_{d=4}~d^4\vec k''\right]\notag\\
&\approx&X(0)+\frac{E_ut}{(2\pi)^4}\int_{[0,2\pi]^4} [\Omega_{xu}+B_{sy}\Omega_{yu}\Omega_{xs}+B_{ys}\Omega_{su}\Omega_{xy}]\left[\left(1+B_{ys}\Omega_{ys}\right)d^4\vec k''\right]\notag\\
&\approx &X(0)+\frac{E_ut}{(2\pi)^4}\int_{[0,2\pi]^4}\Omega_{xu}d^4\vec k''+\frac{B_{ys}E_ut}{(2\pi)^4}\int_{[0,2\pi]^4} [\Omega_{xu}\Omega_{ys}-\Omega_{xs}\Omega_{yu}+\Omega_{xy}\Omega_{su}]d^4\vec k''\notag\\
&=&X(0)+\frac{\mathcal C^{xu}_1}{2\pi}E_ut+\frac{\mathcal C_2}{(2\pi)^2}B_{ys}E_ut
\end{eqnarray}
\end{widetext}
The above derivation began by considering all the above three momentum-resolved wavepacket contributions to $X(t)$, and arrived at the final results via repeated applications of Eq.~\ref{2D}. The phase space modification factor $D_{d=4}= \sqrt{1+B_{\mu\nu}\Omega_{\mu\nu}}+\text{higher order}\approx 1+B_{ys}\Omega_{ys}$ is seen to play a crucial role in producing the term containing $\Omega_{ys}$: without it, the integral giving rise to the 2nd Chern number $\mathcal{C}_2$ will not be complete or even symmetric. Note that higher order corrections must disappear in 4D systems due to antisymmetry.

Indeed, we recover the 2nd-order response equation behavior (Eq.~\ref{eq:curr}) derived earlier. While the 1st Chern number term arises simply from the direct Hall response, the 2nd Chern number term arises via a combination of two different second order Hall responses channels, as well as the density of states correction term $B_{ys}\Omega_{ys}$. While the latter contribution cannot have possibly originated from any Hall pumping, its appearance is necessary for constructing the integrand for the 2nd Chern number.

\subsubsection{Further interpretations for unentangled higher-dimensional QH}

In the case of unentangled (factorizable) 4D states that can be expressed as a tensor product of 2D QH states in $x$-$s$ and $y$-$u$ space, only the product $\Omega_{xs}\Omega_{yu}$ in the integrand of $\mathcal C_2$ survives. 
In this case, the Wannier center flow subject to an electric field $E_u$ and magnetic field $B_0=B_{sy}$ is
\begin{equation}
X_{untangled}^{B_0}(t)=P_X(k_u(t))+P_X(B_{sy}P_Y(k_u(t)))
\label{pumping}
\end{equation}
where $k_u=E_ut$ by minimal coupling, and $P_X$, $P_Y$ are the expressions for Wannier polarization in the $X$ and $Y$ directions. To isolate the 2nd Chern number term from Eq. \ref{pumping}, one can take differences between separate measurements to obtain
\begin{equation}
X_{untangled}^{B_0}(t)-X_{untangled}^{-B_0}(t)=2P_X(B_{sy}P_Y(k_u(t)))
\label{pumping2}
\end{equation}
For the 6D entangled case, one can analogously consider an electric field $E_v$ viz. $k_v=E_vt$, and magnetic fields $B_1=B_{sy}$, $B_2=B_{uz}$. The response of the averaged wavepacket due to all 1st, 2nd and 3rd Chern numbers is thus given by
\begin{eqnarray}
X_{untangled}^{B_1,B_2}(t)&=&P_X(k_v(t))+P_X(B_{sy}P_Y(k_v(t)))\notag\\
&&+P_X(B_{sy}P_Y(B_{uz}P_Z(k_v(t)))
\label{pumping3}
\end{eqnarray}
Similarly, we can take differences and obtain
\begin{eqnarray}
X_{untangled}^{B_1,B_2}(t)-X_{untangled}^{B_1,-B_2}(t)&=&2P_X(B_{sy}P_Y(B_{uz}P_Z(k_v(t)))\notag\\
&\sim& \frac{2C_3}{(2\pi)^3}B_1B_2E_v\,t,
\label{pumping4}
\end{eqnarray}
the final linear approximation holding for large $k_v(t)=E_vt$.

\subsection{Non-commutativity  and phase-space modification}

From the above discussions, it has been apparent that electronic transport does not just arise from ``bare'' electron dynamics, but is also affected by how the phase space distorts around it. This distortion can be interpreted as the ``smudging'' of the electron itself. Promoting the Poisson brackets in Sect.~\ref{sec:DOS} to operator commutators, we see that
\begin{eqnarray}
\omega^{\mu\nu}&=&
-i\left(\begin{matrix}
[\hat r^\mu,\hat r^\nu ]& [\hat k^\mu, \hat r^\nu ] \\
 [\hat r^\mu, \hat k^\nu ] & [\hat k^\mu,\hat k^\nu ]
\end{matrix}\right)\notag\\
&=&\omega_{\mu\nu}^{-1}=\left(\begin{matrix}
  -B & -\mathbb{I}_{d\times d}  \\
\mathbb{I}_{d\times d} & \Omega &
\end{matrix}\right)^{-1}\notag\\
&=&\left(\begin{matrix}
  \Omega(\mathbb{I}-B\Omega)^{-1} & (\mathbb{I}-\Omega B)^{-1} & \\
-(\mathbb{I}-B\Omega)^{-1} &  -(\mathbb{I}-B\Omega)^{-1}B & \end{matrix}\right)\notag\\
&\approx &\left(\begin{matrix}
  \Omega+\Omega B\Omega + ... & \mathbb{I}+\Omega B+(\Omega B)^2+... & \\
-(\mathbb{I}+B\Omega+(B\Omega )^2+...) &  -(B+B\Omega B +...) & \end{matrix}\right)\notag\\
\label{comm}
\end{eqnarray}
To leading order, we simply obtain $[\hat r^\mu,\hat r^\nu ]=i\Omega^{\mu\nu}$, $[\hat k_\mu,\hat k_\nu ]=-iB_{\mu\nu}$ and $[\hat r^\mu,\hat k_\nu]=i(\delta^\mu_\nu+\Omega^{\mu\lambda}B_{\lambda\nu})$, which sets a lower bound for the amount of uncertainty in the phase-space coordinates of an electron, i.e. how much it is ``smudged'' due to the magnetic and Berry curvature fields. As expected, a magnetic field $B$ causes the electrons to move in space cyclotron orbits, thereby smudging it in momentum-space. Analogously, a Berry flux $\Omega$ smudges it in real-space, contributing to a topological obstruction for localized Wannier functions. Nontrivial corrections to these occur precisely when \emph{both} $B$ and $\Omega$ are present. In particular, there will be smudging across the real and momentum space coordinates, with electrons effectively executing phase-space ``cyclotron'' motion that leads to corrections to the transport.

Due to the antisymmetry of both $B$ and $\Omega$ tensors, it can be shown that $\omega^{\mu\nu}$ and hence all commutators are the quotient of the sum of products of matrix elements of $B$ and $\Omega$ with $D(\bs r,\bs k)=\text{Pf}(\omega_{\mu\nu})=\sqrt{\text{Det}(\mathbb{I}-B\Omega)}$. For instance,
\begin{equation}
[\hat r^\mu,\hat k_\mu]=i[(\mathbb{I}-\Omega B)^{-1}]^\mu_\mu=\frac{D_\mu(\bs r,\bs k)}{D(\bs r,\bs k)}
\end{equation}
where $D_\mu$ is the density of states factor in the space with dimension $\mu$ omitted. In general, the commutators have rather complicated expressions in arbitrary number of dimensions involving the explicit form of $(\mathbb{I}-\Omega B)^{-1}$ or $(\mathbb{I}-B\Omega)^{-1}$. Closed form exact expressions can be easily derived for low dimensions however. For $d=2$ dimensions,
\begin{align}
[\hat r^\mu,\hat r^\nu]_{d=2}&=\frac{i\Omega^{\mu\nu}}{1+\Omega^{12}B_{12}}\\
[\hat k_\mu,\hat k_\nu]_{d=2}&=\frac{-iB_{\mu\nu}}{1+\Omega^{12}B_{12}}\\
[\hat r^\mu,\hat k_\nu]_{d=2}&=\frac{i\delta^\mu_\nu}{1+\Omega^{12}B_{12}}
\end{align}
as appearing in Refs.~\cite{xiao2005berry} and \cite{duval2006berry}. For $d=4$ dimensions, $[(\mathbb{I}-B\Omega)^{-1}]^\nu_{\mu}=\frac1{D_{d=4}(\bs r,\bs k)}[\delta^{\nu}_\mu(1-B_{\alpha\beta}\Omega^{\alpha\beta})+B_{\mu\alpha}\Omega^{\alpha\nu}]$, and we have
\begin{align}
[\hat r^\mu,\hat r^\nu]_{d=4}&=i\frac{\Omega^{\mu\nu}(1-B_{\alpha\beta}\Omega^{\alpha\beta})+(\Omega B \Omega)^{\mu\nu}}{D_{d=4}(\bs r,\bs k)}\\
[\hat k_\mu,\hat k_\nu]_{d=4}&=-i\frac{B_{\mu\nu}(1-B_{\alpha\beta}\Omega^{\alpha\beta})+(B\Omega B )_{\mu\nu}}{D_{d=4}(\bs r,\bs k)}\\
[\hat k_\mu,\hat r^\nu]_{d=4}&=\frac{i[\delta^{\nu}_\mu(1-B_{\alpha\beta}\Omega^{\alpha\beta})+B_{\mu\alpha}\Omega^{\alpha\nu}]}{D_{d=4}(\bs r,\bs k)}
\end{align}
where $D_{d=4}(\bs r,\bs k)$ is given in Eq.~\ref{D4D}. These exact expressions should be contrasted with those in the last line of Eq.~\ref{comm}, which do not contain phase space correction denominators.
\begin{figure}[h!]
\centering
\includegraphics[width=.8\linewidth]{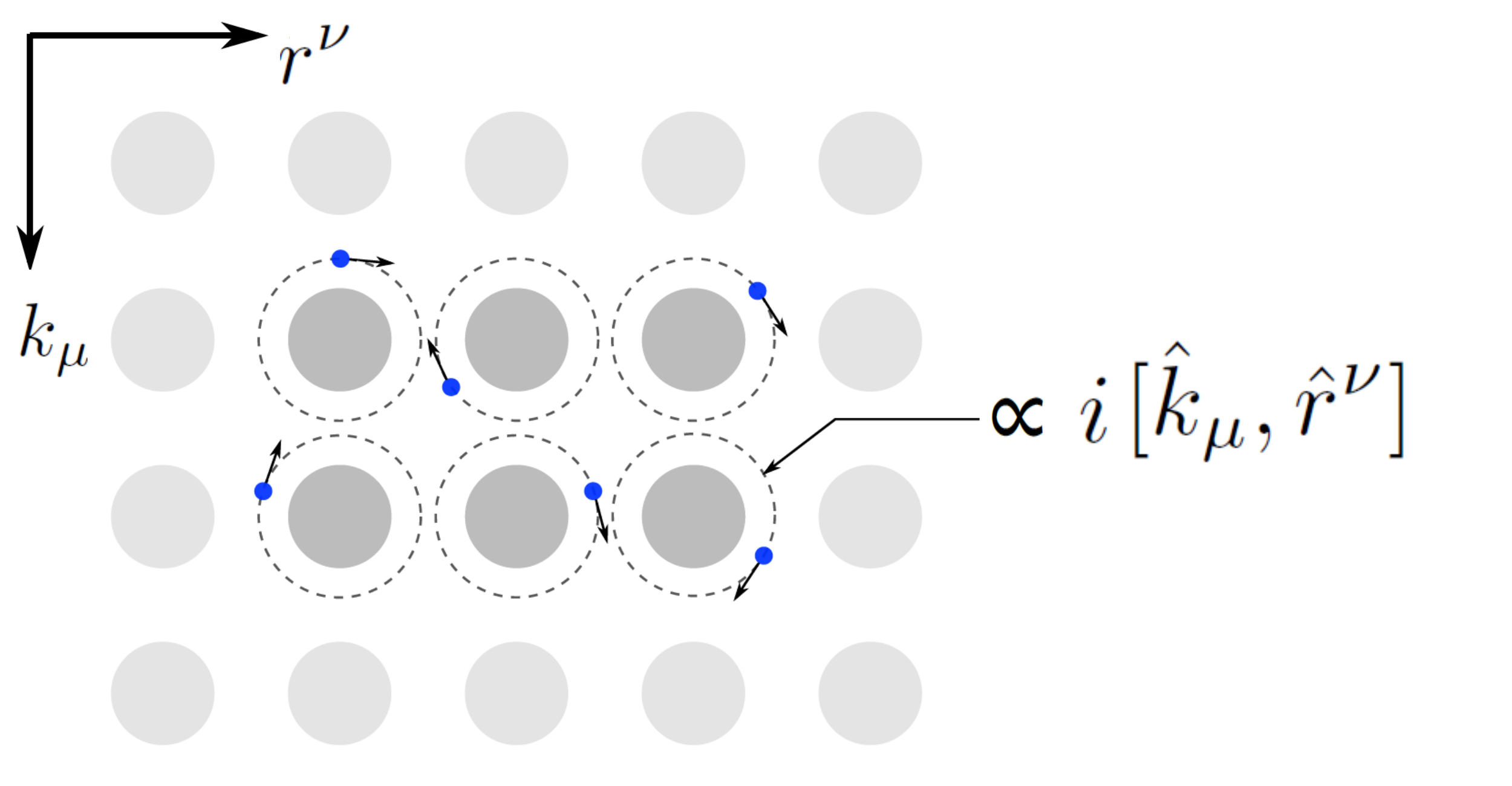}
\caption{Cartoon picture of electrons moving under the simultaneous influence of magnetic field and Berry curvature. They exhibit ``cylotron" orbits in phase space, thereby behaving like ``smudged" particles with uncertainties in their coordinates proportional to the commutator between their phase space coordinates.  }
\end{figure}

\section{Minimal models for novel surface current $j^\mu_{surf}$ in 5D and 6D}
\label{sec:jnon}
In this section, we examine the properties of the new non-topological surface currents through some minimal models with a Fermi surface. These are the simplest models that yield a nonvanishing 2nd order $j^\mu_{surf}$, and can in principle be realized with any tight-binding model by placing nonzero flux across the stipulated plaquettes. Note that we will only be focusing on the current contributions of 2nd order in $B$ and $\Omega$, i.e. second and third lines of Eq.~\ref{jnon}, which appear only in 5 dimensions and beyond.

\subsection{5D case}
Consider a 5D electronic metallic system consisting of 3 spatial dimensions $\{x,y,z\}$ and 2 synthetic dimensions $\{s,w\}$. One possible minimal model consists of just two nonzero components of the magnetic field $B_{\mu\nu}$, and three nonzero components of the Berry curvature $\Omega^{\mu\nu}$. Namely, we have uniform magnetic fields
\begin{align}
B_{xy} & =-B_{yx}  \ne 0,\nonumber \\
B_{sw} & =-B_{ws}  \ne 0,
\end{align}
and uniform Berry curvatures
\begin{align}
\Omega^{xz} & =-\Omega^{zx}  \ne 0,\nonumber \\
\Omega^{ys} & =-\Omega^{sy}\ne 0,\nonumber \\
\Omega^{zw} & =-\Omega^{wz} \ne 0,
\end{align}
This configuration gives a non-topological current density  $j_{surf,5D}^{\mu}$ in the ${\bs e}^z$ direction:
\begin{align}
& j_{surf,5D}^{x} = j_{surf,5D}^{y} =j_{surf,5D}^{s} =j_{surf,5D}^{w} =0 \nonumber \\
j_{surf,5D}^{z}= & B_{xy}B_{sw}\int_{\Gamma} \frac{ \text{d}^6 k }{(2\pi)^6}  \Omega^{sy}\left( \frac{\partial \mathcal{E}}{\partial k_w }\Omega^{xz} + \frac{\partial \mathcal{E}}{\partial k_x }\Omega^{zw}\right)
\end{align}
We see that $j_{surf,5D}^{z}$ should exist for $\Gamma\neq \mathbb{T}^6$ as long as the Berry curvatures are not all uniform, which is generically the case. Note, though that the gradients of $\Omega$, if not already zero, have to be related by the Bianchi identities (see Ref.~\onlinecite{lee2017band} for a systematic way of making the Berry curvatures uniform). $B_{xs}$ and $\Omega^{xz}$ can be implemented as external and intrinsic  physical fluxes, while $B_{yw}, \Omega^{ys}$ and $\Omega^{zw}$ can be implemented as phase factors associated with synthetic parameters $y$ and $w$. Realistic systems will in general possess other non-vanishing fluxes that change the direction of $j_{surf,5D}^{z}$, but they will not cause the latter to vanish when the Fermi surface vanishes.

\subsection{6D case}
Analogous to the 5D case, a minimal model for the 6D metallic case with 3 spatial dimensions $\{x,y,z\}$ and 3 synthetic dimensions $\{s,w,t\}$ is given by uniform magnetic fields
\begin{align}
B_{xw} & =-B_{wx}  \ne 0,\nonumber \\
B_{st} & =-B_{ts}  \ne 0,
\end{align}
and uniform Berry curvatures
\begin{align}
\Omega^{xy} & =-\Omega^{yx}(k_x, k_s)  \ne 0,\nonumber \\
\Omega^{sw} & =-\Omega^{ws}(k_y, k_w)  \ne 0,\\
\Omega^{zt} & =-\Omega^{tz}(k_z, k_t) \ne 0,\nonumber
\end{align}
which gives rise to a non-topological surface current contribution
\begin{align}
& j_{surf,6D}^{x} = j_{surf,6D}^{s} =j_{surf,6D}^{w} =j_{surf,6D}^{t} =0 \nonumber \\
j_{surf,6D}^{z} = & B_{xw}B_{st}\int_{\mathbb{T}^6} \frac{ \text{d}^6 k }{(2\pi)^6} \left(\frac{\partial \mathcal{E}}{\partial k_x } \Omega^{zt} \Omega^{sw}\right) \\
j_{surf,6D}^{y} = & B_{xw}B_{st}\int_{\mathbb{T}^6} \frac{ \text{d}^6 k }{(2\pi)^6} \left(\frac{\partial \mathcal{E}}{\partial k_t } \Omega^{xy} \Omega^{sw}\right)
\end{align}
Both nonzero current components are almost identical, except for the swapping of the roles of $x$ and $t$ components. As before, $\bs j_{surf,6D}$ should not vanish unless the Fermi surface vanish.


\section{Realizations of 6D QH states in 3 physical dimensions} \label{3CN}
Having developed the response formalism for higher-dimensional QH systems, we now present a general approach for realizing them in 3-dimensional physical space.

Building going on to 6D QH models that are realized in 3-dimensional space, we first describe how 2D QH models can be realized in one physical dimension plus one synthetic dimension. In Refs.~\cite{verbin2013observation,madsen2013topological,AAH,4DphotonicQH}, it was shown that a 2D lattice system under the influence of a magnetic field (Hofstadter model) can, with suitable gauge choice, be expressed as a so-called 1D Aubry-Andr\'{e}-Harper (AAH) model with a synthetic parameter. To understand this, consider a square lattice with horizontal ($x$-direction) hoppings $t$ and vertical hoppings $\lambda$. If we choose a gauge such that each vertical hopping (but not horizontal hopping) contains a gauge phase of $e^{2\pi i b x}$, the entire lattice will be threaded with a uniform (internal) magnetic field with each square plaquette experiencing a flux of $2\pi b$. Due to translational symmetry in the vertical direction, we can perform a partial Fourier transform such that $k$ represents the lattice momentum in the vertical direction. The hopping Hamiltonian hence looks like
\begin{align}
H_{2D}=&\sum_k\sum_x\lambda\left(e^{2\pi i bx}e^{ik}+c.c.\right)|\psi_x(k)\rangle\langle \psi_x(k)|\notag\\
&+t\left[|\psi_x(k)\rangle\langle \psi_{x+1}(k)|+|\psi_{x+1}(k)\rangle\langle \psi_{x}(k)|\right]\notag\\
\rightarrow&\sum_{k,x}\left(\lambda\cos(2\pi bx +k)|\psi_x\rangle\langle \psi_x|+t|\psi_x\rangle\langle \psi_{x+1}|+h.c.\right)\notag\\
=&\sum_k H_{1D}^{AAH}(k)
\end{align}
which can be interpreted as a collection of decoupled 1D AAH chains given in the penultimate line. Each AAH chain has an onsite energy term with periodicity $1/b$ and offset $k$, and a uniform nearest-neighbor hopping $t$. For a rational flux $b=\frac{p}{q}$ with $\text{GCD}(p,q)=1$, the AAH Hamiltonian splits into an effective $q$ band Hamiltonian. In this sense, a change of magnetic field, no matter how small, changes the band structure non-perturbatively. Indeed, when the flux $b$ becomes irrational, $H_{1D}^{AAH}$ becomes a quasiperiodic chain with dispersionless Landau Levels. That said, the physical responses remain perfectly smoothly defined, as derived in the previous section where the fields are taken as external perturbations.

\subsection{Unentangled 6D QH system with 3rd Chern number}
\begin{figure}[h!]
\centering
\includegraphics[width=.85\linewidth]{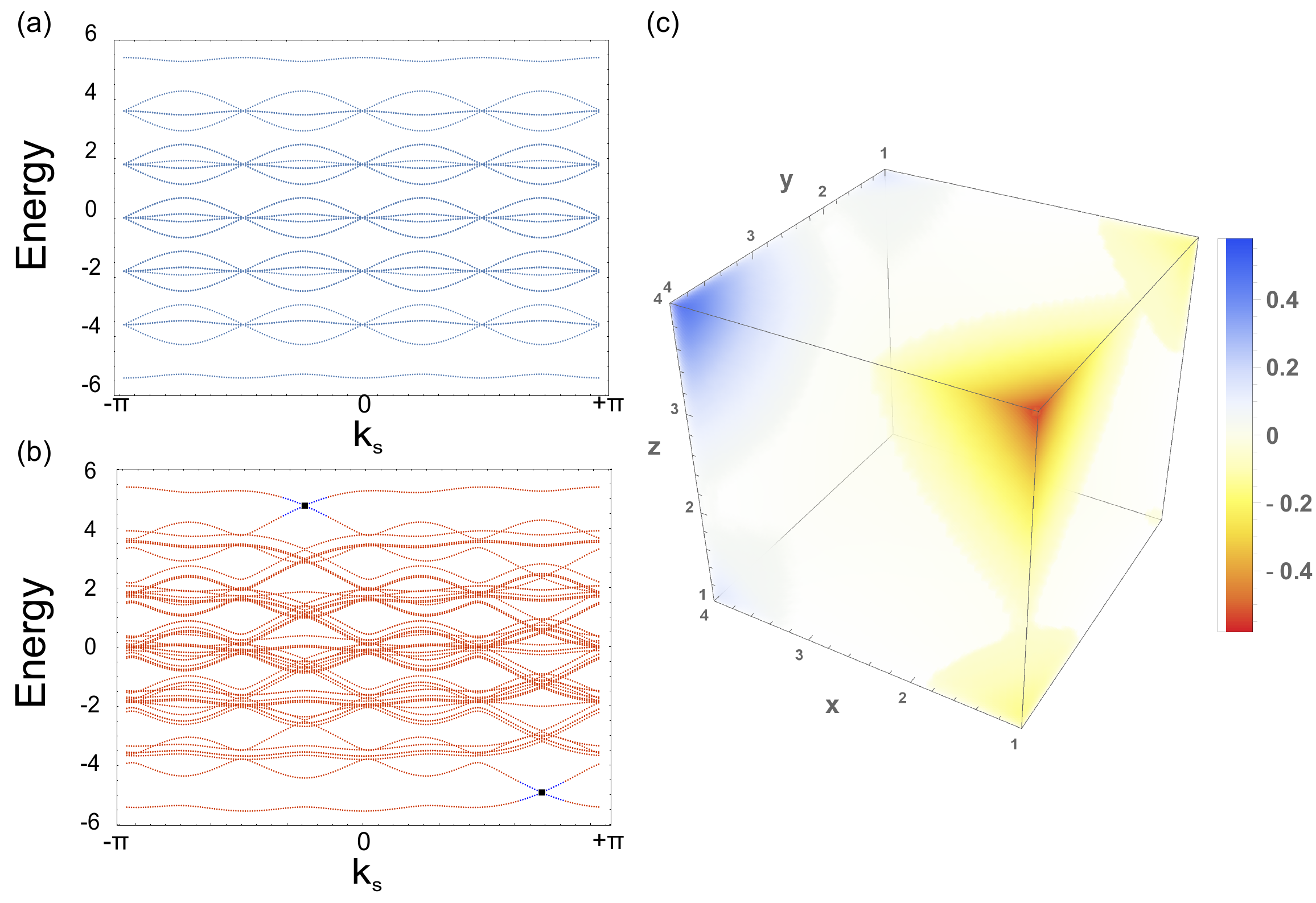}
\caption{a) The energy spectrum of the 6D AAH model (Eq.~\ref{TB1}) with $\lambda_1=\lambda_2=\lambda_3=3t=1.5$ and $b=1/4$ numerically solved with a $4\times 4\times 4 $ super-cell for a) periodic and b) open boundary conditions, where gapless states (small black boxes) that traverse the bulk gap appear, indicative of a nontrivial Chern number. 
c) The density plot of the energy eigenstate (small black box at $k_w=k_t=k_x=k_y=k_z=0, k_s=-0.783$) for open boundaries in the x-direction. We observe clear localization in the x-surface, in fact at its corners due to $C_4$ symmetry.}
\label{fig:AAH1}
\end{figure}

The most direct way of achieving a 6D QH system is to construct it as a tensor product of three 2D QH systems, each represented by a 1D AAH model, i.e.
\begin{align}
H_{6D}=&\sum_{\bs x}\left[\sum_{i}\lambda_i\cos(2\pi bx_i +k_i)\right]|\psi_{x_1,x_2,x_3}\rangle\langle \psi_{x_1,x_2,x_3}|\notag\\
&+t\left(|\psi_{x_1+1,x_2,x_3}\rangle\langle\psi_{x_1,x_2,x_3}| +h.c.\,\right)\notag\\
&+t\left(|\psi_{x_1,x_2+1,x_3}\rangle\langle\psi_{x_1,x_2,x_3}| +h.c.\,\right)\notag\\
&+t\left(|\psi_{x_1,x_2,x_3+1}\rangle\langle\psi_{x_1,x_2,x_3}| +h.c.\,\right)
\label{TB1}
\end{align}
whose energy spectrum is shown in Fig.~\ref{fig:AAH1} for flux $b=p/q=1/4$, i.e. with each magnetic unit cell consisting of $q=4$ sites.


As a tensor product of three AAH models, this 6D QH state possess a 3rd Chern number~\cite{qi2008topological} that is the tensor product of three 1st Chern numbers:
\begin{align}
\mathcal{C}_3=&\frac1{3!(2\pi)^3}\int_{\mathbb{T}^6} \Omega\wedge \Omega\wedge \Omega\notag\\
=&\frac1{(2\pi)^3}\left(\int_{\mathbb{T}^2}\Omega \right)\times \left(\int_{\mathbb{T}^2}\Omega \right)\times \left(\int_{\mathbb{T}^2}\Omega \right)\notag\\
=&\mathcal{C}_1^{xs}\mathcal{C}_1^{yw}\mathcal{C}_1^{zt},
\label{C3}
\end{align}
with the 3 spatial dimensions and their 3 corresponding synthetic dimensions labeled as $x,y,z$ and $s,w,t$ respectively.



\subsection{Entangled 6D QH system with 3rd Chern number}
\label{sec:entangled}
Although the 6D system described by Eq.~\ref{TB1} is a bona-fide 6D QH system with electronic response given by the earlier sections, its spectral flow properties can be entirely understood in terms of those of its constituent 1st Chern number systems. To generate a 3rd Chern number system that is not such a trivial tensor product, the simplest approach is to mix (or \emph{entangle}) their directions of spatial modulation~\cite{4DphotonicQH}: $(x_1,x_2,x_3)^T\rightarrow R(x_1,x_2,x_3)^T$, where $R$ is a orthogonal matrix. Since the $t$-couplings that define the 3D lattice remain unchanged, the $R$ rotation effectively couples the different copies of the AAH models. If the $R$ matrix is replaced by a generic non-degenerate 3-by-3 matrix, we will obtain a 6D QH system that is mixed in a nontrivial though not necessarily symmetrical manner.

\begin{figure}[h!]
\centering
\includegraphics[width=.85\linewidth]{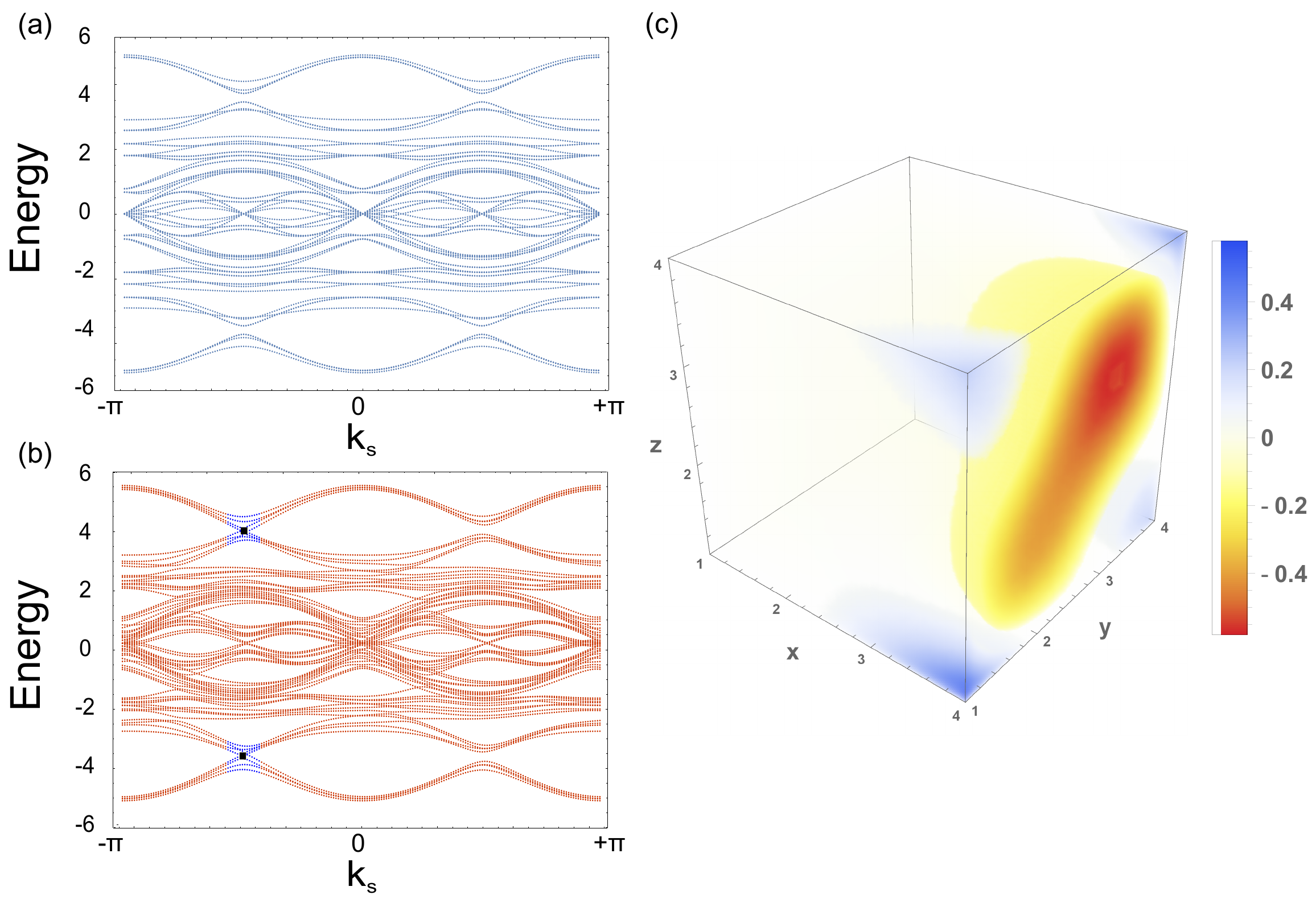}
\caption{The energy spectrum of the entangled 6D AAH model (Eq.~\ref{TB2}), also with $\lambda_=\lambda_2=\lambda_3=3t=1.5$ and $b=1/4$, for a) periodic and b) open boundary conditions. Due to the nontrivial coupling between the different copies of AAH models, numerous additional sub-bands appear. Gapless degenerate topological boundary states (small black boxes) that traverse the bulk gap appear, and are similarly plotted in c) as in the previous figure except that $k_s = -1.553$. The state is clearly localized in the x-surface. }
\label{fig:AAH2}
\end{figure}

For illustration, consider the rotated onsite energy modulation of the form
\begin{align}
E(x,y,z)= & \cos(2\pi b (x+y-z)+k_s)\nonumber\\
& + \cos(2\pi b (x-y+z)+k_w) \nonumber \\
& + \cos(2\pi b (y+z-x)+k_t),
\label{eq2}
\end{align}
with $b$ rescaled for simplicity. The original  synthetic dimension parameters $k_s$, $k_w$ and $k_t$ remain unchanged. Explicitly, we obtain the entangled 6D QH Hamiltonian
\begin{align}
H'_{6D}=&\lambda \sum_{x,y,z}E(x,y,z)|\psi_{x,y,z}\rangle\langle \psi_{x,y,z}|\notag\\
&+t\left(|\psi_{x+1,y,z}\rangle\langle\psi_{x,y,z}| +h.c.\,\right)\notag\\
&+t\left(|\psi_{x,y+1,z}\rangle\langle\psi_{x,y,z}| +h.c.\,\right)\notag\\
&+t\left(|\psi_{x,y,z+1}\rangle\langle\psi_{x,y,z}| +h.c.\,\right)
\label{TB2}
\end{align}
whose spectrum is plotted in Fig.~\ref{fig:AAH2}.


\section{Physical realization and topological response measurement of 3rd Chern number}

In the previous section, we have discussed how periodically modulated AAH systems with synthetic dimensional offsets can possess higher Chern numbers. In the quasiperiodic limit, the Chern bands become flat Landau levels, effectively simulating higher-dimensional QH systems with uniform magnetic field. It is important to note, however, that quasiperiodicity is \emph{not} a necessary ingredient - Chern bands can emerge even with a relatively small unit cell, as long as the modulations are defined to result in nonvanishing effective flux.

\begin{figure}[h!]
\centering
\includegraphics[width=\linewidth]{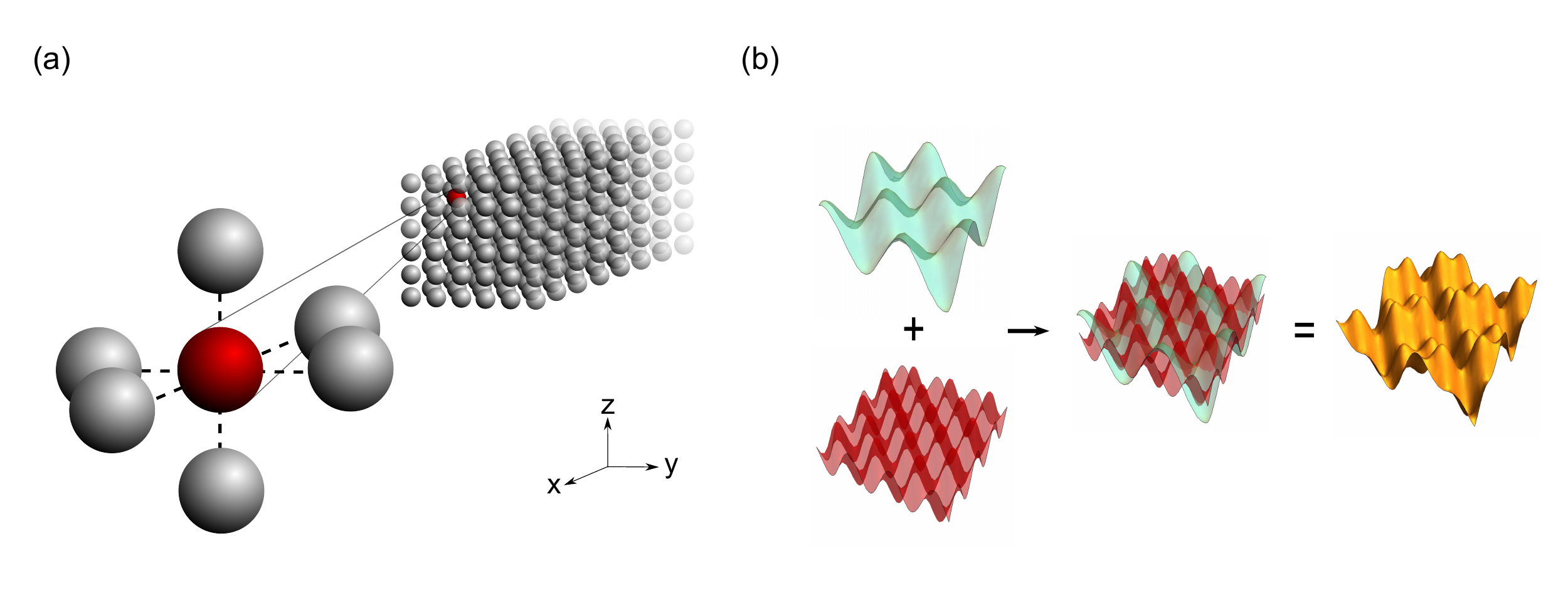}
\caption{The cold-atom setup for the 6D quantum Hall (QH) system of Eq.~\ref{Vs2}, with external magnetic fields $B_{xw}$ and $B_{yt}$, illustrated for the unentangled case $R=\mathbb{I}_{3\times 3}$.
a) 3D cartoon of the cold-atom system. The lattice sites are formed by the potential in Fig. \ref {fig:exp}(b), with each sphere denoting a cold atom.
b) The system is modulated periodically by the sum of two sub-potentials (orange), as given by Eq.~\ref{Vs2}. The blue potential denotes the on-site modulation we set to the system and the red potential denotes the potential that pins the cold atoms to the lattice sites.}
\label{fig:exp}
\end{figure}

For realistic implementation, we consider a 3D optical lattice of ultracold atoms, the 2D version for which has already been experimentally realized~\cite{Celi2014Synthetic,Lohse2017Exploring}. 
Consider an optical potential that is modulated by two sinusoidal profiles in each direction, as illustrated in Fig.~\ref{fig:exp}:
\begin{equation}
\Phi(\bs x)=\sum_i \left(\phi_{i}\sin^2 (b\pi x_i) +\phi'_{i}\sin^2 (b'\pi (Rx_i)-k_i)\right)
\label{Vs2}
\end{equation}
where $R$ is the rotation matrix introduced in Sect.~\ref{sec:entangled}, $b,b'$ are dissimilar inverse periods and $\phi_i,\phi'_i$ are modulation potential depths which can be different in different directions $i$. Due to the interference between the two modulations, we obtain an effective flux that leads to higher Chern bands, as elaborated in Ref.~\cite{Lohse2017Exploring}. When $R\neq \mathbb{I}_{3\times 3}$, i.e. in Eqs.~\ref{eq2} or \ref{TB2}, the constituent Chern systems are coupled, resulting in an entangled higher Chern (or QH) phase. The synthetic momenta are implemented by the phase offsets $k_i$, which can be controlled by slightly tilting the optical lattice.

For simplicity, $1/b$ can be taken to be the spacing of the 3D lattice (Fig.~\ref{fig:exp}b), while $1/b'$ shall correspond to an imposed modulation period (Fig.~\ref{fig:exp}c). The potential $\phi_i$ is given by the blue potential in Fig.\ref{fig:exp}c, which fixes the atoms on lattice sites, while the potential $\phi'_i$ introduces the local modulation as given by the red potential. Taking the s-wave approximation, Eq.~\ref{Vs2} describes a spatially modulated tight-binding Hamiltonian similar to Eq.~\ref{TB2}, with effective flux of magnitude $\frac{2\pi b\hbar}{eaa'}$ in a direction determined by the matrix $R$, where $e$ is the electronic charge and $a,a'$ are the lattice constants in the two directions connected by $R$.

To detect the higher-dimensional QH transport properties, an external electric field $E_\mu$ has to be introduced into the cold-atom system. For that, one can implement a linear gradient~\cite{Brantut2012Conduction} either magnetically~\cite{PhysRevLett.111.185301,Miyake2013Realizing} or optically~\cite{Aidelsburger2014Measuring}. The current, whether $\bs j_{surf}$ or the topological contribution, can also be indirectly measured via the center-of-mass velocity of an atomic cloud after accounting for the density of states correction~\cite{Dauphin2013Extracting,Atala2014Observation}.

For illustration, we focus on the lowest 6D band, which for appropriate choices of uniform magnetic fluxes is non-degenerate and well-isolated from the higher bands. Suppose all the Berry curvature components are zero except for:
\begin{align}
\Omega^{xs} & =-\Omega^{sx}(k_x, k_s) \ne 0, \nonumber \\
\Omega^{yw} & =-\Omega^{wy}(k_y, k_w)  \ne 0,\\
\Omega^{zt} & =-\Omega^{tz}(k_x, k_s) \ne 0, \nonumber
 \label{eq:non-zero}
\end{align}
We will choose, with no loss of generality, the perturbing (synthetic) electric field to be along the $s$ direction i.e. $\bs E\!=\!E_s \bs e^z$ as in Fig.\ref{fig:exp} a. However, in 6D, different choices for the perturbing magnetic field will lead to dramatic differences in the observables. 
If we choose the only nonzero external magnetic field components to be $B_{xw}$ and $B_{yt}$, the current density ${\bs j}$ in this experimental setup is explicitly given by:
\begin{align}
j^{x} =&  \mathcal C_1 E_{s} \,, \nonumber \\
j^{y} =& -\frac{\mathcal C_{2}}{(2 \pi)^4} E_{s} B_{xw} \,,  \nonumber \\
j^{z} =& \frac{\mathcal C_{3}}{(2 \pi)^3} E_{s} B_{xw} B_{yt} \,, \nonumber \\
j^{s} =& j^{w} = j^{t}=0
\end{align}
which is schematically shown in Fig.\ref{fig:exp}a. The various Chern numbers $\mathcal C_1,\mathcal C_2$ and $\mathcal C_3$ can then be independently obtained by measuring the various current density components.

\section{Conclusion}
Starting from the very first principles, we have derived electromagnetic response properties of arbitrarily-dimensioned QH systems through the complementary approaches of semiclassical wavepacket dynamics and topological pumping arguments. Together, they illustrate the crucial role of a modified phase space density of states when magnetic field and Berry curvature are simultaneously present. This modification becomes all the more nontrivial in higher dimensions, where it provide the crucially missing pieces in the Chern invariants for the topological response. We derived some new mathematical results for it and interpreted them in terms of non-commuting phase space dynamics.

Notably, we have explored new surface current contributions $\bs j_{surf}$ in dimensions greater than four, which appear in addition to the usual topological terms. We provided minimal models for investigating this enigmatic term, at least up to its leading order. These models, as well as the paradigmatic 6D QH systems, can be realized in realistic 3D experimental setups like cold atom systems through possibly entangled products of AAH chains.

Subsequent to the initial appearance of our manuscript on arXiv, we become aware of a related work (Ref.~\onlinecite{petrides20186d}) on higher-dimensional QH systems, which provided a more in-depth description of experimental realization using three-dimensional topological charge pumps in cold atomic systems.

\section*{ACKNOWLEDGMENTS}
C.H.L. and Y.Z.W. contributed equally to this work. We thank Oded Zilberberg and Hannah Price for useful discussions. X.Z. is supported by the National Natural Science Foundation of China (Grant No.11404413), the Natural Science Foundation of Guangdong Province (Grant No. 2015A030313188) and the Guangdong
Science and Technology Innovation Youth Talent Program (Grant No. 2016TQ03X688).
\bibliography{apssamp}
\newpage
\clearpage
\section*{Appendices}

\setcounter{section}{0}

\section{Method for computing the determinant of a special type of antisymmetric matrix}
\label{sec:det}
\begin{widetext}
In deriving Eq.~\ref{Drk} from the Pfaffian of Eq.~\ref{omegamunu} in the main text, we need to find the determinant of matrices of the form
\begin{align}
M=
\left(
\begin{array}{cccccccccc}
    0 & -X_{a_{1}b_{1}} & -X_{a_{1}b_{2}} & \cdots & -X_{a_{1}b_{k}} & -1 & 0 & 0 & \cdots & 0\\
    X_{a_{1}b_{1}} & 0 & -X_{a_{2}b_{2}} & \cdots & -X_{a_{2}b_{k}} & 0 & -1 & 0 & \cdots & 0\\
    X_{a_{1}b_{2}} & X_{a_{2}b_{2}} & 0 & \cdots & -X_{a_{3}b_{k}} & 0 & 0 & -1 & \cdots & 0\\
    \vdots & \vdots & \vdots & \ddots & \vdots & \vdots & \vdots & \vdots & \ddots & \vdots\\
    X_{a_{1}b_{k}} & X_{a_{2}b_{k}} & X_{a_{3}b_{k}} & \cdots & 0 & 0 & 0 & 0 & \cdots & -1\\
    1 & 0 & 0 & \cdots & 0 & 0 & Y_{a_{1}b_{1}} & Y_{a_{1}b_{2}} & \cdots & Y_{a_{1}b_{k}}\\
    0 & 1 & 0 & \cdots & 0 & -Y_{a_{1}b_{1}} & 0 & Y_{a_{2}b_{2}} & \cdots & Y_{a_{2}b_{k}}\\
    0 & 0 & 1 & \cdots & 0 & -Y_{a_{1}b_{2}} & -Y_{a_{2}b_{2}} & 0 & \cdots & Y_{a_{3}b_{k}}\\
    \vdots & \vdots & \vdots & \ddots & \vdots & \vdots & \vdots & \vdots & \ddots & \vdots\\
    0 & 0 & 0 & \cdots & 1 & -Y_{a_{1}b_{k}} & -Y_{a_{2}b_{k}} & -Y_{a_{3}b_{k}} & \cdots & 0
\end{array}
\right).
\label{M}
\end{align}
\end{widetext}
We will adopt a mathematical induction approach for this computation, and exploit the the antisymmetry of $M$ and the symmetry between the indices of $X_{a_{i}b_{j}}$ and $Y_{a_{i}b_{j}}$. First, we give the conditions that the expansion of the determinant should satisfy:\\
\\
\begin{itshape}

$\circ$ The expansion should be a perfect square.

$\circ$ $X_{a_{i}b_{j}}$ and $Y_{a_{i}b_{j}}$ should be symmetric and non-interfering.

$\circ$ There should not be any odd-ordered term in the expansion.

$\circ$ 
Suppose the power of the expansion is $m$, and the dimensionality of the matrix is $n=2k$ ($k$ is a nonnegative integer,because obviously $n$ is even), then
\begin{center}
\[
m=\left\{
\begin{array}{cc}
2(k-1), & \mbox{k is a nonnegative odd integer;}\\
2k, & \mbox{k is a nonnegative even integer.}
\end{array}
\right.
\]
\end{center}
\end{itshape}

The first condition is based on the antisymmetry of the matrix. As we know, the expansion of any antisymmetric determinant must be a square of a polynomial~\cite{CayleySur} which is called the Pfaffian of the matrix. And it's apparent to get the second condition by observing the characteristics of this matrix, which proclaim that $X_{a_{i}b_{j}}$ and $Y_{a_{i}b_{j}}$ are symmetric to some extent. The expansion of the determinant stays the same in the case of exchanging all the $X$ and $Y$ with same indexes. Changing any elements of the determinant will not influence the form of the other elements in the expansion. As for the third condition, there has to be one zero at least in the odd-ordered terms, and thus they will not exist in the expansion. The last condition are satisfied when $k$ is even because apparently the power of the expansion equals the dimensionality of the matrix. To discuss the odd-$k$ situation, we need to find out why the highest order term vanishes.

Since the highest order term of the determinant will not contain $1,-1$ or $0$ factors in its determinant expansion, we can deduce that it is decided by these elements:
\begin{widetext}
\begin{align}
\left(
\begin{array}{cccccccccc}
      & -X_{a_{1}b_{1}} &  &  & -X_{a_{1}b_{k}} &  &  &  &  & \\
    X_{a_{1}b_{1}} &  & -X_{a_{2}b_{2}} &  &  &  &  &  &  & \\
     & X_{a_{2}b_{2}} &  & \ddots &  &  &  &  &  & \\
     &  & \ddots &  & - X_{a_{k}b_{k}} &  &  &  &  & \\
    X_{a_{1}b_{k}} &  &  & X_{a_{k}b_{k}} &  &  &  &  &  & \\
     &  &  &  &  &  & Y_{a_{1}b_{1}} &  &  & Y_{a_{1}b_{k}}\\
     &  &  &  &  & -Y_{a_{1}b_{1}} &  & Y_{a_{2}b_{2}} &  & \\
     &  &  &  &  &  & -Y_{a_{2}b_{2}} &  & \ddots & \\
     &  &  &  &  &  & \ddots & &  & Y_{a_{k}b_{k}}\\
     &  &  &  &  &  -Y_{a_{1}b_{k}} &  & -Y_{a_{k}b_{k}} &  &
\end{array}
\right)
\end{align}
\end{widetext}

Hence we can find that there are just four ways to form a highest order term. However, considering the definition of determinant:
\begin{align}
det(M)=\varepsilon_{i_{1} \cdot \cdot \cdot i_{n} } a_{1,i_{1}} a_{2,i_{2}} \cdot \cdot \cdot a_{n,i_{n}}
\end{align}
where $\varepsilon_{i_{1} \cdot \cdot \cdot i_{n} }$ is a Levi-Civita symbol, we can get these four highest order terms $h_1,h_2,h_3$ and $h_4$, which have the same absolute values:
\begin{align}
|h_1|& =|h_2|=|h_3|=|h_4| \nonumber \\
& =\resizebox{.85\hsize}{!}{$\displaystyle{|X_{a_{1}b_{k}}(X_{a_{1}b_{1}}X_{a_{2}b_{2}} \cdots X_{a_{k}b_{k}})Y_{a_{1}b_{k}}(Y_{a_{1}b_{1}}Y_{a_{2}b_{2}} \cdots Y_{a_{k}b_{k}})|}$}
\end{align}
and when $k$ is even, these four terms are all nonnegative; When $k$ is odd, two of these four terms are nonnegative which are the opposite number of the other two terms. Thus it's obvious that these four terms cancel each other out, which causes the power of the matrices with an odd $k$ decreases accordingly.

Before giving the conjectured solution, we need to introduce the concept of a general Levi-Civita symbol, whose indices don't have to be a permutation consisted of continuous natural number.
With
 \begin{align}
\alpha_{1}, \alpha_{2}, \alpha_{3}, \cdots \alpha_{n}  \in \mathds{N_{+}} \nonumber
\end{align}
and $p$ the parity of the permutation in the above sequence,
\begin{align}
\varepsilon_{\alpha_{1} \alpha_{2} \alpha_{3} \cdots \alpha_{n}}=\left\{
\begin{array}{cc}
1, & \mbox{$p$ is even ;}\\
-1, & \mbox{$p$ is odd;}\\
0, & \mbox{if any two indices are equal.}\\
\end{array}
\right.
\end{align}
As we have established the four conditions that the expansion should satisfy, we shall now put forward the conjecture solution:
\begin{widetext}
\begin{align}
&\sqrt{\mbox{det} (M)} \notag\\
& =  1 + \frac{1}{2}  X_{ \alpha_{11} \beta_{11} } Y_{ \alpha_{11} \beta_{11} }   + \frac{1}{ 2^{6} } ( \varepsilon_{ \alpha_{21} \beta_{21} \alpha_{22} \beta_{22} }   X_{ \alpha_{21} \beta_{21} } X_{ \alpha_{22} \beta_{22} } ) \times ( \varepsilon_{ \alpha_{23} \beta_{23} \alpha_{24} \beta_{24} } Y_{ \alpha_{23} \beta_{23} } Y_{ \alpha_{24} \beta_{24} } ) \nonumber \\
& + \cdot \cdot \cdot \cdot\cdot\cdot\cdot\cdot\cdot\cdot\nonumber \\
&+ \frac{1}{ 2^{4\times\lfloor\frac{n}{4}\rfloor-2}} ( \varepsilon_{ \alpha_{(\lfloor\frac{n}{4}\rfloor, 1)} \beta_{(\lfloor\frac{n}{4}\rfloor, 1)} \alpha_{(\lfloor\frac{n}{4}\rfloor, 2)} \beta_{(\lfloor\frac{n}{4}\rfloor, 2)} \cdot \cdot \cdot  \alpha_{(\lfloor\frac{n}{4}\rfloor, \lfloor\frac{n}{4}\rfloor) } \beta_{(\lfloor\frac{n}{4}\rfloor, \lfloor\frac{n}{4}\rfloor) }} \nonumber X_{\alpha_{(\lfloor\frac{n}{4}\rfloor,1)} \beta_{(\lfloor\frac{n}{4}\rfloor, 1) }} X_{ \alpha_{(\lfloor\frac{n}{4}\rfloor, 2)} \beta_{(\lfloor\frac{n}{4}\rfloor, 2)} } \cdot \cdot \cdot X_{ \alpha_{(\lfloor\frac{n}{4}\rfloor, \lfloor\frac{n}{4}\rfloor)} \beta_{(\lfloor\frac{n}{4}\rfloor, \lfloor\frac{n}{4}\rfloor) }}  ) \nonumber\\
& \times( \varepsilon_{ \alpha_{(\lfloor\frac{n}{4}\rfloor, \lfloor\frac{n}{4}\rfloor+1)} \beta_{(\lfloor\frac{n}{4}\rfloor, \lfloor\frac{n}{4}\rfloor+1)} \alpha_{(\lfloor\frac{n}{4}\rfloor, \lfloor\frac{n}{4}\rfloor+2)} \beta_{(\lfloor\frac{n}{4}\rfloor, \lfloor\frac{n}{4}\rfloor+2)} \cdot \cdot \cdot  \alpha_{(\lfloor\frac{n}{4}\rfloor, 2\lfloor\frac{n}{4})\rfloor } \beta_{（\lfloor\frac{n}{4}\rfloor, 2\lfloor\frac{n}{4}\rfloor) }}  \nonumber\\
&Y_{\alpha_{(\lfloor\frac{n}{4}\rfloor, \lfloor\frac{n}{4}\rfloor+1)} \beta_{(\lfloor\frac{n}{4}\rfloor, \lfloor\frac{n}{4}\rfloor+1) }} Y_{ \alpha_{(\lfloor\frac{n}{4}\rfloor, \lfloor\frac{n}{4}\rfloor+2)} \beta_{(\lfloor\frac{n}{4}\rfloor, \lfloor\frac{n}{4}\rfloor+2)} } \cdot \cdot \cdot Y_{ \alpha_{(\lfloor\frac{n}{4}\rfloor, 2\lfloor\frac{n}{4})\rfloor} \beta_{(\lfloor\frac{n}{4}\rfloor, 2\lfloor\frac{n}{4}\rfloor) }}  ) \\
& + \cdot \cdot \cdot \cdot\cdot\cdot\cdot\cdot\cdot\cdot \,,\nonumber
\end{align}
\end{widetext}

where $n$ is the dimensionality of the original matrix $M$ and the set that consists of the indices of $X$ equals the one that consists of the indices of $Y$ in the same term. The reason why we leave an ellipsis behind the whole expansion is that in fact, there exist infinite terms in the expansion but the terms behind all vanish due to the definition of Levi-Civita symbol, which means we only need to calculate this formula to the order that we require, and  $\alpha_{ij}$ and $\beta_{ij}$  satisfy:
\begin{align}
\alpha_{ij} \in \{a_{k}\}, \beta_{ij} \in \{b_{k}\}
\end{align}

Moreover,the elements of the matrix $X_{a_{i}b_{j}} $  satisfy:
\begin{align}
X_{a_{i}b_{j}}=-X_{b_{j}a_{i}}
\end{align}
and $f(x)=\lfloor x \rfloor$ is the floor function.

Since we have allowed the dimensionality of the matrix to be any arbitrarily large number, we shall employ mathematical induction to prove that this expansion is exactly the right one.

First, when $n=1,2,3$, it's apparent that the expansion equals the determinants of the matrices involved. Provided that the proposition is valid when $n=k$, for $n=k+1$ we have:
\begin{widetext}
\begin{align}
det(M)=
& \left|
\begin{array}{cccccccccc}
    0 & -X_{a_{1}b_{1}} & -X_{a_{1}b_{2}} & \cdots & -X_{a_{1}b_{k}} & -1 & 0 & 0 & \cdots & 0\\
    X_{a_{1}b_{1}} & 0 & -X_{a_{2}b_{2}} & \cdots & -X_{a_{2}b_{k}} & 0 & -1 & 0 & \cdots & 0\\
    X_{a_{1}b_{2}} & X_{a_{2}b_{2}} & 0 & \cdots & -X_{a_{3}b_{k}} & 0 & 0 & -1 & \cdots & 0\\
    \vdots & \vdots & \vdots & \ddots & \vdots & \vdots & \vdots & \vdots & \ddots & \vdots\\
    X_{a_{1}b_{k}} & X_{a_{2}b_{k}} & X_{a_{3}b_{k}} & \cdots & 0 & 0 & 0 & 0 & \cdots & -1\\
    1 & 0 & 0 & \cdots & 0 & 0 & Y_{a_{1}b_{1}} & Y_{a_{1}b_{2}} & \cdots & Y_{a_{1}b_{k}}\\
    0 & 1 & 0 & \cdots & 0 & -Y_{a_{1}b_{1}} & 0 & Y_{a_{2}b_{2}} & \cdots & Y_{a_{2}b_{k}}\\
    0 & 0 & 1 & \cdots & 0 & -Y_{a_{1}b_{2}} & -Y_{a_{2}b_{2}} & 0 & \cdots & Y_{a_{3}b_{k}}\\
    \vdots & \vdots & \vdots & \ddots & \vdots & \vdots & \vdots & \vdots & \ddots & \vdots\\
    0 & 0 & 0 & \cdots & 1 & -Y_{a_{1}b_{k}} & -Y_{a_{2}b_{k}} & -Y_{a_{3}b_{k}} & \cdots & 0
\end{array}
\right| \nonumber \\
=
& \left|
\begin{array}{cccccccccccc}
    0 & -X_{a_{1}b_{1}} & -X_{a_{1}b_{2}} & \cdots & -X_{a_{1}b_{k}} & 0  & -1 & 0 & 0 & \cdots & 0 & 0\\
    X_{a_{1}b_{1}} & 0 & -X_{a_{2}b_{2}} & \cdots & -X_{a_{2}b_{k}} & 0 & 0 & -1 & 0 & \cdots & 0 & 0 \\
    X_{a_{1}b_{2}} & X_{a_{2}b_{2}} & 0 & \cdots & -X_{a_{3}b_{k}} & 0 & 0 & 0 & -1 & \cdots & 0 & 0 \\
    \vdots & \vdots & \vdots & \ddots & \vdots & \vdots & \vdots & \vdots & \vdots & \ddots & \vdots & \vdots \\
    X_{a_{1}b_{k}} & X_{a_{2}b_{k}} & X_{a_{3}b_{k}} & \cdots & 0 & 0 & 0 & 0 & 0 & \cdots & -1 & 0 \\
    0 & 0 & 0 & \cdots & 0 & 0 & 0 & 0 & 0 & \cdots & 0 & -1 \\
    1 & 0 & 0 & \cdots & 0 & 0 & 0 & Y_{a_{1}b_{1}} & Y_{a_{1}b_{2}} & \cdots & Y_{a_{1}b_{k}}& 0 \\
    0 & 1 & 0 & \cdots & 0 & 0 & -Y_{a_{1}b_{1}} & 0 & Y_{a_{2}b_{2}} & \cdots & Y_{a_{2}b_{k}}& 0 \\
    0 & 0 & 1 & \cdots & 0 & 0 & -Y_{a_{1}b_{2}} & -Y_{a_{2}b_{2}} & 0 & \cdots & Y_{a_{3}b_{k}}& 0 \\
    \vdots & \vdots & \vdots & \ddots & \vdots & \vdots & \vdots & \vdots & \vdots & \ddots & \vdots& \vdots \\
    0 & 0 & 0 & \cdots & 1 & 0 & -Y_{a_{1}b_{k}} & -Y_{a_{2}b_{k}} & -Y_{a_{3}b_{k}} & \cdots & 0 & 0 \\
    0 & 0 & 0 & \cdots & 0 & 1 & 0 & 0 & 0 & \cdots & 0 & 0
\end{array}
\right|
\end{align}
\end{widetext}

which shows that when:
\begin{align}
X_{a_{i}b_{k+1}}  = & Y_{a_{j} b_{k+1}} =0\\
i,j  \in \mathds{Z}_{+} & ,i \leq k+1, j \leq  k+1 \nonumber
\end{align}
then:
\begin{align}
det(M_{2k})=det(M_{2(k+1)})
\end{align}
where $M_{n}$ is a $n$-dimensional antisymmetric matrix derived from zero-padding $M$ when necessary. Apparently, the determinants of $M_n$ and $M$ are equal. Hence we only need to consider the extra terms that these new elements in the matrix bring to the determinant. With that, the expansion of $det(M_{2(k+1)})$ must be of the form:
\begin{align}
\sqrt{\mbox{det} (M_{2k})} =\sqrt{\mbox{det} (M_{2(k+1)})} & + P(X_{a_{i}b_{k+1}}, Y_{a_{j} b_{k+1}})\\
i,j  \in \mathds{Z}_{+} & ,i \leq k+1, j \leq  k+1 \nonumber
\end{align}

where $P(X_{a_{i}b_{k+1}}, Y_{a_{j} b_{k+1}})$ is the polynomial that contains $X_{a_{i}b_{k+1}}$ and $Y_{a_{j} b_{k+1}}$ ($i,j  \in \mathds{Z}_{+}, i \leq k+1, j \leq k+1 $).

Combined with the four rules we mentioned above, we can calculate the $k+1$-th order determinant in this way: first we keep the absolute values of all the new $Y$ terms be $1$,and whether the specific term is positive or negative depends on its location in the determinant. In our case, the terms above the diagonal terms are positive and the terms below the diagonal terms are negative, which is like:
\begin{widetext}
\begin{align}
\left|
 \begin{array}{cccccccccccc}
     0 & -X_{a_{1}b_{1}} & -X_{a_{1}b_{2}} & \cdots & -X_{a_{1}b_{k}} & -X_{a_{1}b_{k+1}}  & -1 & 0 & 0 & \cdots & 0 & 0\\
     X_{a_{1}b_{1}} & 0 & -X_{a_{2}b_{2}} & \cdots & -X_{a_{2}b_{k}} & -X_{a_{1}b_{k
     +1}} & 0 & -1 & 0 & \cdots & 0 & 0 \\
     X_{a_{1}b_{2}} & X_{a_{2}b_{2}} & 0 & \cdots & -X_{a_{3}b_{k}} & -X_{a_{1}b_{k+1}} & 0 & 0 & -1 & \cdots & 0 & 0 \\
     \vdots & \vdots & \vdots & \ddots & \vdots & \vdots & \vdots & \vdots & \vdots & \ddots & \vdots & \vdots \\
     X_{a_{1}b_{k}} & X_{a_{2}b_{k}} & X_{a_{3}b_{k}} & \cdots & 0 & -X_{a_{1}b_{k+1}} & 0 & 0 & 0 & \cdots & -1 & 0 \\
     X_{a_{1}b_{k+1}} & X_{a_{1}b_{k+1}} & X_{a_{1}b_{k+1}} & \cdots & X_{a_{1}b_{k+1}} & 0 & 0 & 0 & 0 & \cdots & 0 & -1 \\
     1 & 0 & 0 & \cdots & 0 & 0 & 0 & Y_{a_{1}b_{1}} & Y_{a_{1}b_{2}} & \cdots & Y_{a_{1}b_{k}}& 1 \\
     0 & 1 & 0 & \cdots & 0 & 0 & -Y_{a_{1}b_{1}} & 0 & Y_{a_{2}b_{2}} & \cdots & Y_{a_{2}b_{k}}& 1 \\
     0 & 0 & 1 & \cdots & 0 & 0 & -Y_{a_{1}b_{2}} & -Y_{a_{2}b_{2}} & 0 & \cdots & Y_{a_{3}b_{k}}& 1 \\
     \vdots & \vdots & \vdots & \ddots & \vdots & \vdots & \vdots & \vdots & \vdots & \ddots & \vdots& \vdots \\
     0 & 0 & 0 & \cdots & 1 & 0 & -Y_{a_{1}b_{k}} & -Y_{a_{2}b_{k}} & -Y_{a_{3}b_{k}} & \cdots & 0 & 1 \\
     0 & 0 & 0 & \cdots & 0 & 1 & -1 & -1 & -1 & \cdots & -1 & 0
 \end{array}
 \right|
 \end{align}
 \end{widetext}
Then we can expand the determinant according to the definition. We know that the result we get is not the general one, which is not easy to get. However considering the symmetry between $X$ terms and $Y$ terms, we can get the general result by adjust the formula artificially, which means that we complement the formula to a symmetric one. For instance, if we have got a term that contains $X_{a_{1}b_{k+1}}$, which might as well be named $T_{1,k+1}$, there must exists a term that contains exactly the same elements as $T_{1,k+1}$ except changing $X_{a_{1}b_{k+1}}$ into $Y_{a_{1}b_{k+1}}$. Hence we can get:
\begin{widetext}
\begin{align}
P(X_{a_{i}b_{k+1}}, Y_{a_{j} b_{k+1}}) = &  \frac{1}{2}  X_{ \alpha_{11} \beta_{11} } Y_{ \alpha_{11} \beta_{11} }   + \frac{1}{ 2^{6} } ( \varepsilon_{ \alpha_{21} \beta_{21} \alpha_{22} \beta_{22} }   X_{ \alpha_{21} \beta_{21} } X_{ \alpha_{22} \beta_{22} } ) \times ( \varepsilon_{ \alpha_{23} \beta_{23} \alpha_{24} \beta_{24} } Y_{ \alpha_{23} \beta_{23} } Y_{ \alpha_{24} \beta_{24} } ) \nonumber \\
& + \cdot \cdot \cdot \cdot\cdot\cdot\cdot\cdot\cdot\cdot\nonumber \\
&+  \resizebox{.85\hsize}{!}{$\displaystyle{\frac{1}{ 2^{4\times\lfloor\frac{k+1}{4}\rfloor-2}} ( \varepsilon_{ \alpha_{(\lfloor\frac{k+1}{4}\rfloor, 1)} \beta_{(\lfloor\frac{k+1}{4}\rfloor, 1)} \alpha_{(\lfloor\frac{k+1}{4}\rfloor, 2)} \beta_{(\lfloor\frac{k+1}{4}\rfloor, 2)} \cdot \cdot \cdot  \alpha_{(\lfloor\frac{k+1}{4}\rfloor, \lfloor\frac{k+1}{4}\rfloor) } \beta_{(\lfloor\frac{k+1}{4}\rfloor, \lfloor\frac{k+1}{4}\rfloor) }} \nonumber X_{\alpha_{(\lfloor\frac{k+1}{4}\rfloor,1)} \beta_{(\lfloor\frac{k+1}{4}\rfloor, 1) }} X_{ \alpha_{(\lfloor\frac{k+1}{4}\rfloor, 2)} \beta_{(\lfloor\frac{k+1}{4}\rfloor, 2)} } \cdot \cdot \cdot X_{ \alpha_{(\lfloor\frac{k+1}{4}\rfloor, \lfloor\frac{k+1}{4}\rfloor)} \beta_{(\lfloor\frac{k+1}{4}\rfloor, \lfloor\frac{k+1}{4}\rfloor) }}  )}$} \nonumber\\
& \times( \varepsilon_{ \alpha_{(\lfloor\frac{k+1}{4}\rfloor, \lfloor\frac{k+1}{4}\rfloor+1)} \beta_{(\lfloor\frac{k+1}{4}\rfloor, \lfloor\frac{k+1}{4}\rfloor+1)} \alpha_{(\lfloor\frac{k+1}{4}\rfloor, \lfloor\frac{k+1}{4}\rfloor+2)} \beta_{(\lfloor\frac{k+1}{4}\rfloor, \lfloor\frac{k+1}{4}\rfloor+2)} \cdot \cdot \cdot  \alpha_{(\lfloor\frac{k+1}{4}\rfloor, 2\lfloor\frac{k+1}{4})\rfloor } \beta_{（\lfloor\frac{k+1}{4}\rfloor, 2\lfloor\frac{k+1}{4}\rfloor) }}  \nonumber\\
&Y_{\alpha_{(\lfloor\frac{k+1}{4}\rfloor, \lfloor\frac{k+1}{4}\rfloor+1)} \beta_{(\lfloor\frac{k+1}{4}\rfloor, \lfloor\frac{k+1}{4}\rfloor+1) }} Y_{ \alpha_{(\lfloor\frac{k+1}{4}\rfloor, \lfloor\frac{k+1}{4}\rfloor+2)} \beta_{(\lfloor\frac{k+1}{4}\rfloor, \lfloor\frac{k+1}{4}\rfloor+2)} } \cdot \cdot \cdot Y_{ \alpha_{(\lfloor\frac{k+1}{4}\rfloor, 2\lfloor\frac{k+1}{4})\rfloor} \beta_{(\lfloor\frac{k+1}{4}\rfloor, 2\lfloor\frac{k+1}{4}\rfloor) }}  )  \,,\nonumber
\end{align}
\end{widetext}
where one of the indices of each term must be the additional $(k+1) th$ index. Thus we can prove the correctness of our conjectured solution. Back to the problem of the modified phase-space density of states, we just need to consider a matrix of the form
\begin{widetext}
\begin{align}
 \left(
\begin{array} {cccccccccccc}
 0 & - B_{x y} & - B_{x s} & - B_{x z} & - B_{x w} & - B_{x t} & -1 & 0 & 0 & 0 & 0 & 0\\
 B_{x y} & 0 & - B_{y s} & - B_{y z} & - B_{y w} & - B_{y t} & 0 & -1 & 0 & 0 & 0 & 0\\
 B_{x s} & B_{y s} & 0 & - B_{s z} & - B_{s w} & - B_{s t} & 0 & 0 & -1 & 0 & 0 & 0\\
 B_{x z} & B_{y z} & B_{s z} & 0 & - B_{z w} & - B_{z t} & 0 & 0 & 0 & -1 & 0 & 0\\
 B_{x w} & B_{y w} & B_{s w} & B_{z w} & 0 & - B_{w t} & 0 & 0 & 0 & 0 & -1 & 0\\
 B_{x t} & B_{y t} & B_{s t} & B_{z t} & B_{w t} & 0 & 0 & 0 & 0 & 0 & 0 & -1\\
 1 & 0 & 0 & 0 & 0 & 0 & 0 & \Omega^{x y} & \Omega^{x s} & \Omega^{x z} & \Omega^{x w} & \Omega^{x t}\\
 0 & 1 & 0 & 0 & 0 & 0 & - \Omega^{x y} & 0 & \Omega^{y s} & \Omega^{y z} & \Omega^{y w} & \Omega^{y t}\\
 0 & 0 & 1 & 0 & 0 & 0 & - \Omega^{x s} & - \Omega^{y s} & 0 & \Omega^{s z} & \Omega^{s w} & \Omega^{s t}\\
 0 & 0 & 0 & 1 & 0 & 0 & - \Omega^{x z} & - \Omega^{y z} & - \Omega^{s z} & 0 & \Omega^{z w} & \Omega^{z t}\\
 0 & 0 & 0 & 0 & 1 & 0 & - \Omega^{x w} & - \Omega^{y w} & - \Omega^{s w} & - \Omega^{z w} & 0 & \Omega^{w t}\\
 0 & 0 & 0 & 0 & 0 & 1 & - \Omega^{x t} & - \Omega^{y t} & - \Omega^{s t} & - \Omega^{z t} & - \Omega^{w t} & 0\\
   \end{array}
\right)
 \left(
\begin{array} {c}
\dot{x} \\
\dot{y} \\
\dot{s} \\
\dot{z} \\
\dot{w} \\
\dot{t} \\
\dot{k}_x \\
\dot{k}_y \\
\dot{k}_s \\
\dot{k}_z \\
\dot{k}_w\\
\dot{k}_t
\end{array}
\right)
=
 \left(
\begin{array} {c}
E_x \\
E_y \\
E_s \\
E_z \\
E_w \\
E_t \\
\frac{\partial \mathcal{E}}{\partial k_x }\\
\frac{\partial \mathcal{E}}{\partial k_y } \\
\frac{\partial \mathcal{E}}{\partial k_s }\\
\frac{\partial \mathcal{E}}{\partial k_z } \\
\frac{\partial \mathcal{E}}{\partial k_w }\\
\frac{\partial \mathcal{E}}{\partial k_t }
\end{array}
\right)\,,
\end{align}
\end{widetext}
which we now know must have a Pfaffian given by Eq.~\ref{Drk}.
%
\\
\newpage
\section{Derivation of 6D Current Density and Transport Equations  from Third-Order Semi-classics}
\label{sec:derivation}
Here, we fill in the details of the derivation of Eq.~\ref{eq:curr}, specializing to $D=6$ dimensions for concreteness. To calculate the response current density of a filled band, we set the integration (phase space) measure to be $ \int_{\mathbb{T}^6} \text{d}^6 k \,D({\bs r}, {\bs k})$, where $D({\bs r}, {\bs k}) $ is the modified phase-space density of states, which we have derived in the previous Appendix. In 6D, it is given by%
\begin{align}
D({\bs r}, {\bs k}) = & \frac{1}{(2\pi)^6} \left[  1 + \frac{1}{2}  B_{\mu \nu} \Omega^{\mu \nu}  \right. +\frac{1}{64} \left( \varepsilon^{\alpha \beta \gamma \delta } B_{\alpha \beta }B_{\gamma \delta }\right)  \nonumber \\
&  \times \left( \varepsilon_{\mu \nu \lambda \rho} {\Omega}^{\mu \nu }\Omega^{\lambda \rho }\right) + \frac{1}{1024} \left( \varepsilon^{\zeta \eta \theta \tau \kappa \xi } B_{\zeta \eta }B_{\theta \tau }B_{\kappa \xi }  \right) \nonumber \\
&\times  \left. \left( \varepsilon_{\sigma \omega \iota \phi \chi \psi } {\Omega}^{\sigma \omega }{\Omega}^{\iota \phi }{\Omega}^{\chi \psi }\right) \right] \,, \label{eq:dos}
\end{align}
with $\mu\!=\!x,y,z,s,w,t$. Combining this expression with EOMs Eqs.~\ref{eq:semir} and \ref{eq:semik} from the main text, we obtain, up to third order,
\,\label{eq:current}
 \begin{widetext}
\begin{align}
j^\mu =& \int_{\Gamma} { \text{d}^6 k } [ \dot{{ r}}^\mu D({\bs r}, {\bs k})]  \label{eq:intcurrent} \\
 \approx& \int_{\Gamma} \frac{ \text{d}^6 k }{(2\pi)^6}
  \left[ v_\mu  +  E_\nu  \Omega^{\mu \nu} +
\left(  E_\delta \Omega^{\gamma \delta}B_{ \nu \gamma} \Omega^{\mu \nu} +  \frac{1}{2}E_\nu  \Omega^{\mu \nu} B_{\delta \gamma}  \Omega^{\delta \gamma}\right)
+
  \left( v_\gamma   B_{ \nu \gamma} \Omega^{\mu \nu} + \frac{1}{2} v_\mu B_{\gamma \nu}  \Omega^{\gamma \nu} \right)
   \right.
  \nonumber \\
&+ \left( \left( v_\alpha  B_{ \delta \alpha} \Omega^{\gamma \delta}
+ \frac{1}{2} v_\gamma   B_{\delta \alpha}  \Omega^{\delta \alpha}  \right) B_{ \nu \gamma} \Omega^{\mu \nu}+
  \frac{1}{64}  v_\mu ( \varepsilon^{\alpha \beta \gamma \delta } B_{\alpha \beta }B_{\gamma \delta} )( \varepsilon_{\xi \nu \lambda \rho} \Omega^{\xi \nu }\Omega^{ \lambda \rho} )   \right) \nonumber \\
&+ \left( \frac{1}{64}  E_\nu  \Omega^{\mu \nu} \left( \varepsilon^{\alpha \beta \gamma \delta } B_{\alpha \beta }B_{\gamma \delta }\right) \times \left( \varepsilon_{\theta \nu \lambda \rho} {\Omega}^{\theta \nu }\Omega^{\lambda \rho }\right)  + \frac{1}{2} E_\delta  \Omega^{\gamma \delta}B_{ \nu \gamma}  \Omega^{\mu \nu}  B_{\beta \theta} \Omega^{\beta \theta} \nonumber + E_\beta \Omega^{\alpha \beta} B_{ \delta \alpha}  \Omega^{\gamma \delta} B_{ \nu \gamma}  \Omega^{\mu \nu} \right)\\
&+ \left( \frac{1}{1024} v_\mu \left( \varepsilon^{\zeta \eta \theta \tau \kappa \xi } B_{\zeta \eta }B_{\theta \tau }B_{\kappa \xi }  \right) \times  \left( \varepsilon_{\sigma \omega \iota \phi \chi \psi } {\Omega}^{\sigma \omega }{\Omega}^{\iota \phi }{\Omega}^{\chi \psi }\right)+ \frac{1}{64} v_\gamma  B_{ \nu \gamma} \Omega^{\mu \nu}  v_\mu \left( \varepsilon^{\alpha \beta \gamma \delta } B_{\alpha \beta }B_{\gamma \delta} \right)\times \left( \varepsilon_{\xi \nu \lambda \rho} \Omega^{\xi \nu }\Omega^{ \lambda \rho} \right) \right. \nonumber \\
&+ \left. \left. \frac{1}{2}  B_{\beta \theta} \Omega^{\beta \theta} v_\alpha  B_{ \delta \alpha} \Omega^{\gamma \delta} B_{ \nu \gamma}  \Omega^{\mu \nu} + v_\theta  B_{ \beta \theta} \Omega^{\alpha \beta}  B_{ \delta \alpha}  \Omega^{\gamma \delta} B_{ \nu \gamma}  \Omega^{\mu \nu} \right)\right]
\,,\label{eq:current}
 \end{align}
 \end{widetext}
The first term disappears unless the Fermi surface $\partial\Gamma$ is non-vanishing, because $\int_\Gamma d^6k \,v_\mu$ is a total differential. The second term simply gives the 1st Chern number response. The next set of terms can be shown to mostly cancel off due to antisymmetry of both the magnetic field and the Berry curvature, except for 2nd Chern number response term:
\begin{align}
&\int_{\Gamma} \frac{ \text{d}^6 k }{(2\pi)^6}
\left(   E_\delta \Omega^{\gamma \delta}B_{ \nu \gamma} \Omega^{\mu \nu} +  \frac{1}{2}E_\nu  {\Omega}^{\mu \nu} B_{\delta \gamma}  {\Omega}^{\delta \gamma}\right) \notag\\
&\rightarrow \frac{1}{2} \frac{\mathcal C_2}{(2  \pi)^4} \varepsilon^{\mu \alpha \beta \nu} E_\nu B_{ \alpha \beta} \, ,
\end{align}
if $\Gamma=\mathbb{T}^6$. The following set of terms turns out to evaluate to zero by virtue of Bianchi's identity~\cite{Xiao2009Polarization}, unless there is a non-vanishing Fermi surface $\partial \Gamma$:
\begin{align}
&\int_{\Gamma} \frac{ \text{d}^6 k }{(2\pi)^6}    \left(  v_\gamma   B_{ \nu \gamma} \Omega^{\mu \nu} + \frac{1}{2} v_\mu B_{\gamma \nu}  \Omega^{\gamma \nu} \right)  \notag\\
=&  \int_{\Gamma} \frac{ \text{d}^6 k }{(2\pi)^6}
\mathcal{E} \left(
\frac{\partial \Omega^{\nu \gamma} }{\partial k_\mu }+   \frac{\partial \Omega^{ \gamma \mu} }{\partial k_\nu }+  \frac{\partial \Omega^{\mu \nu} }{\partial k_\gamma }  \right)B_{\nu \gamma}  =0
\end{align}
The next set of terms
\begin{align}
 \int_{\Gamma} &\frac{ \text{d}^6 k }{(2\pi)^6}
  \left[ \left( v_\alpha  B_{ \delta \alpha} \Omega^{\gamma \delta}
+ \frac{1}{2}v_\gamma   B_{\delta \alpha}  {\Omega}^{\delta \alpha}  \right) B_{ \nu \gamma} \Omega^{\mu \nu}\right. \nonumber \\
& + \left.
  \frac{1}{64}  v_\mu ( \varepsilon^{\alpha \beta \gamma \delta } B_{\alpha \beta }B_{\gamma \delta} )\times( \varepsilon_{\xi \nu \lambda \rho} {{\Omega}}^{\xi \nu }{\Omega}^{ \lambda \rho} )   \right]  \,.
 \end{align}
can be shown to also disappear unless there is a Fermi surface i.e. $\Gamma\neq \mathbb{T}^6$. We write the integrand as
%
\begin{align}
v_\mu  \times \frac{1}{64} \sum_{i}|\varepsilon^{\mu \alpha \beta \gamma \delta }| \varepsilon^{ \alpha \beta \gamma \delta } B_{\alpha \beta }B_{\gamma \delta } \Omega_{i} ^{\ast} (\alpha \beta \gamma \delta)
\end{align}
where we have defined a new function $\Omega^{\ast}$ for ease of notation:
\begin{small}
\begin{center}
\[
\Omega_{i}^{\ast} \!(\alpha \!, \! \beta \!, \! \gamma \!, \! \delta)=\left\{
\begin{array}{cc}
\varepsilon^{ P_{i 1} \! P_{i 2} \! P_{i 3} \! P_{i 4}}\Omega^{P_{i 1} \! P_{i 2}}\Omega^ {P_{i 3} \! P_{i 4}}, & \mbox{$\varepsilon^{ P_{i 1} \! P_{i 2} \! P_{i 3} \! P_{i 4}}\neq 0$;}  \\
     &     \\
0, & \mbox{ $\varepsilon^{ P_{i 1} \! P_{i 2} \! P_{i 3} \! P_{i 4}}\! = \! 0 $ }
\end{array}
\right.
\]
\end{center}
\end{small}
$P$ is the matrix of permutations of $\alpha, \beta, \gamma$ and $ \delta $, i.e.
\begin{align}
P(\alpha,\beta,\gamma,\delta)=
\left(
\begin{array}{cccc}
  \alpha & \beta & \gamma & \delta\\
  \alpha & \beta & \delta & \gamma\\
  \alpha & \gamma & \beta & \delta\\
  \alpha & \gamma & \delta & \beta\\
  \alpha & \delta & \beta & \gamma\\
  \alpha & \delta & \gamma & \beta\\
  \beta & \alpha & \gamma & \delta\\
  \beta & \alpha & \delta & \gamma\\
  \beta & \gamma & \alpha & \delta\\
  \beta & \gamma & \delta & \alpha\\
  \beta & \delta & \alpha & \gamma\\
  \beta & \delta & \gamma & \alpha\\
  \gamma & \alpha & \beta & \delta\\
  \gamma & \alpha & \delta & \beta\\
  \gamma & \beta & \alpha & \delta\\
  \gamma & \beta & \delta & \alpha\\
  \gamma & \delta & \alpha & \beta\\
  \gamma & \delta & \beta & \alpha\\
  \delta & \alpha & \beta & \gamma\\
  \delta & \alpha & \gamma & \beta\\
  \delta & \beta & \alpha & \gamma\\
  \delta & \beta &\gamma & \alpha\\
  \delta & \gamma & \alpha & \beta\\
  \delta & \gamma & \beta & \alpha\\
\end{array}
\right)
\label{M}
\end{align}
Together, all the terms of this set form the non-topological current contribution $\bs j_{surf}$ in the main text. With the above analysis, one sees that although it has a complicated form, it turns out to be zero under certain circumstances. For instance, if the energy $\mathcal{E}$ is an even function of $k$, each term in the integral will be zero. Also, using the setup in our main text, we find this term turns out to be:
\begin{align}
j_{surf}^{z}= & \int_{\Gamma} \frac{ \text{d}^6 k }{(2\pi)^6} (v_x  B_{xw}B_{yt}\Omega^{st} \Omega^{yw}) \\
j_{surf}^{s}= & \int_{\Gamma} \frac{ \text{d}^6 k }{(2\pi)^6} (v_t  B_{xw}B_{yt}\Omega^{xz} \Omega^{yw})\\
j_{surf}^{x}= & j_{surf}^{y}=j_{surf}^{w}=j_{surf}^{t}=0
\end{align}
where $j_{surf}^{z}$ and $j_{surf}^{s}$ turns out to be zero after integrating, too.

The next set of terms
\begin{align}
\int_{\Gamma} & \frac{ \text{d}^6 k }{(2\pi)^6} \left( \frac{1}{64}  E_\nu  \Omega^{\mu \nu} \left( \varepsilon^{\alpha \beta \gamma \delta } B_{\alpha \beta }B_{\gamma \delta }\right) \times \left( \varepsilon_{\theta \nu \lambda \rho} {\Omega}^{\theta \nu }\Omega^{\lambda \rho }\right)\right.  \nonumber \\
& + \left. \frac{1}{2} E_\delta  \Omega^{\gamma \delta}B_{ \nu \gamma}  \Omega^{\mu \nu}  B_{\beta \theta} \Omega^{\beta \theta} \nonumber + E_\beta \Omega^{\alpha \beta} B_{ \delta \alpha}  \Omega^{\gamma \delta} B_{ \nu \gamma}  \Omega^{\mu \nu} \right)
\end{align}
can also be simplified using the antisymmetry of the magnetic field strength ($B_{ \gamma \nu}= - B_{ \nu \gamma}$) and the Berry curvature ($\Omega^{ \gamma \nu}=-\Omega^{ \nu \gamma}$), and consequently for a vanishing Fermi surface reduces to
\begin{align}
\frac{1}{8} \frac{\mathcal{C}_3}{8 \pi^3}  \varepsilon^{\mu \alpha \beta  \delta \gamma \nu}  E_\nu B_{ \alpha \beta} B_{\delta \gamma}
 \end{align}
where $\mathcal C_3$ is the 3rd Chern number as defined in the main text.
The final group of terms
\begin{align}
\int_{\Gamma} & \frac{ \text{d}^6 k }{(2\pi)^6}
 \left( \resizebox{.85\hsize}{!}{$\displaystyle{\frac{1}{1024} v_\mu \left( \varepsilon^{\zeta \eta \theta \tau \kappa \xi } B_{\zeta \eta }B_{\theta \tau }B_{\kappa \xi }  \right) \times  \left( \varepsilon_{\sigma \omega \iota \phi \chi \psi } {\Omega}^{\sigma \omega }{\Omega}^{\iota \phi }{\Omega}^{\chi \psi }\right)}$} \right. \nonumber \\
 &+ \resizebox{.85\hsize}{!}{$\displaystyle{\frac{1}{64} v_\gamma  B_{ \nu \gamma} \Omega^{\mu \nu}  v_\mu \left( \varepsilon^{\alpha \beta \gamma \delta } B_{\alpha \beta }B_{\gamma \delta} \right)\times \left( \varepsilon_{\xi \nu \lambda \rho} \Omega^{\xi \nu }\Omega^{ \lambda \rho} \right)}$} \nonumber \\
 &+ \left.  \resizebox{.85\hsize}{!}{$\displaystyle{\frac{1}{2}  B_{\beta \theta} \Omega^{\beta \theta} v_\alpha  B_{ \delta \alpha} \Omega^{\gamma \delta} B_{ \nu \gamma}  \Omega^{\mu \nu} + v_\theta   B_{ \beta \theta} \Omega^{\alpha \beta}  B_{ \delta \alpha}  \Omega^{\gamma \delta} B_{ \nu \gamma}  \Omega^{\mu \nu}}$} \right) \,.
\end{align}
fortuitously all cancel to zero by virtue of the antisymmetry of the magnetic field and of the Berry curvature in 6D, reminiscent of the $\bs j_{surf}$ terms which cancel totally in 4D due to the same symmetries. We conjecture that this final set of terms will contribute to a higher $\bs j_{surf}$ in dimensions greater than $6$, although proving that becomes rather tedious. Combining all the above results, we obtain the expressions in Eqs.~\ref{eq:curr} and~\ref{jnon} in the main text.

\end{document}